\documentclass[11pt, a4paper]{article}

\usepackage{amsmath}
\usepackage{amsfonts}
\usepackage{amssymb}
\usepackage{bbm}
\usepackage{verbatim}
\usepackage{dsfont}
\usepackage{booktabs}
\usepackage{slashed}
\usepackage{nicefrac}
\usepackage{longtable}

\usepackage{epsfig}
\usepackage{color}
\usepackage[table]{xcolor}
\usepackage{graphicx}
\usepackage[font=small]{caption}
\usepackage[listofformat=empty,subrefformat=empty]{subfig}

\usepackage{float}
\usepackage{overpic}
\usepackage{tikz}
\usepackage{tikz-3dplot}
\usetikzlibrary{calc}
\usetikzlibrary{decorations} 
\usepackage{tikz-cd} 

\usepackage{bbold}

\usepackage{enumerate}
\usepackage{hhline}
\usepackage{multirow}

\usepackage[nosort]{cite}
\usepackage{xspace}
\usepackage{setspace}

\usepackage[pdftitle={On the Gauge Group Topology of 8d CHL Vacua},
  pdfauthor={Mirjam Cvetic, Markus Dierigl, Ling Lin, Hao Y. Zhang},
  pdfsubject={string landscape, swampland, 8d supergravity, 1-form symmetry},
  bookmarksopen, bookmarksnumbered, bookmarksopenlevel=2, colorlinks=false, linkcolor=blue, citecolor=blue, urlcolor=blue]{hyperref}

\numberwithin{equation}{section}

\newcommand{\bbZ}{\mathbb{Z}}

\newcommand{\fkg}{{\mathfrak{g}}}
\newcommand{\fkh}{{\mathfrak{h}}}
\renewcommand{\tfrac}{\genfrac{}{}{}1}

\tikzstyle{A}=[fill=none, draw=black, shape=circle]
\tikzstyle{none}=[fill=none, draw=none, shape=circle]

\newcommand{\tabincell}[2]{\begin{tabular}{@{}#1@{}}#2\end{tabular}}

\title{On the Gauge Group Topology of 8d CHL Vacua}
\author{Mirjam Cveti\v{c}, Markus Dierigl, Ling Lin, Hao Y. Zhang}
\date{June 2021}

\begin{document}

\baselineskip=18pt

\vspace*{-2cm}
\begin{flushright}
	\texttt{CERN-TH-2021-104} \\
	\texttt{UPR-1313-T}\\
  \texttt{LMU-ASC 19/21}
\end{flushright}

%% title, authors, affiliation
\vspace*{0.6cm} 
\begin{center}
{\LARGE{\textbf{On the Gauge Group Topology of 8d CHL Vacua}}} \\
 \vspace*{1.5cm}
Mirjam Cveti{\v c}$^{1,2,3}$, Markus Dierigl$^4$, Ling Lin$^{5}$, Hao Y.~Zhang$^{1}$\\

{
 \vspace*{1.0cm} 
{\it ${}^1$ Department of Physics and Astronomy, \\University of Pennsylvania,
Philadelphia, PA 19104, USA\\ \vspace{.2cm}
${}^2$ Department of Mathematics, \\University of Pennsylvania,
Philadelphia, PA 19104, USA\\ \vspace{.2cm}
${}^3$ Center for Applied Mathematics and Theoretical Physics,\\
University of Maribor, SI20000 Maribor, Slovenia\\ \vspace{.2cm}
${}^4$ Arnold-Sommerfeld-Center for Theoretical Physics,\\
Ludwig-Maximilians-Universit\"at, 80333 M\"unchen, Germany\\ \vspace{.2cm}
${}^5$CERN Theory Department, CH-1211 Geneva, Switzerland}
}

\vspace*{0.8cm}
\end{center}
\vspace*{.5cm}

\noindent
Compactifications of the CHL string to eight dimensions can be characterized by embeddings of root lattices into the rank 12 momentum lattice $\Lambda_M$, the so-called Mikhailov lattice.
Based on this data, we devise a method to determine the global gauge group structure including all $U(1)$ factors.
The key observation is that, while the physical states correspond to vectors in the momentum lattice, the gauge group topology is encoded in its dual.
Interpreting a non-trivial $\pi_1(G) \equiv {\cal Z}$ for the non-Abelian gauge group $G$ as having gauged a ${\cal Z}$ 1-form symmetry, we also prove that all CHL gauge groups are free of a certain anomaly \cite{Cvetic:2020kuw} that would obstruct this gauging.
We verify this by explicitly computing ${\cal Z}$ for all 8d CHL vacua with rank$(G)=10$.
Since our method applies also to $T^2$ compactifications of heterotic strings, we further establish a map that determines any CHL gauge group topology from that of a ``parent'' heterotic model.

\thispagestyle{empty}
\clearpage

\setcounter{page}{1}

\newpage

\begingroup
  \flushbottom
 \tableofcontents
\endgroup
% \newpage

\vspace*{1cm}

\section{Introduction}

Supersymmetric string compactifications on low-dimensional internal manifolds have seen a resurgence of interest within the Swampland program \cite{Vafa:2005ui,Ooguri:2006in}.
One of the main reasons is that, thanks to the large amount of supersymmetry, one can essentially classify all supergravity models that arise as the low-energy description of such compactifications.
Therefore, they provide an excellent ``laboratory'' to test our understanding of the physical principles that separate the Landscape from the Swampland.

Given the profound role of gauge symmetries in our mathematical formulation of effective theories, principles that delineate the boundary between consistent and inconsistent gauge groups of supergravity models are of particular interest.
In the context of 8d ${\cal N}=1$ supergravity theories, significant progress in this direction has been made recently, which not only explains the absence of specific gauge \emph{algebras} \cite{Garcia-Etxebarria:2017crf,Montero:2020icj,Hamada:2021bbz} in the 8d string landscape, but also some of the intricate patterns of the possible global structures, i.e., topology, of the gauge \emph{group} \cite{Cvetic:2020kuw,Montero:2020icj}.
In particular, the ideas pertaining to the gauge group topology have been mostly tested and confirmed for 8d theories with total gauge rank\footnote{
Within the known 8d ${\cal N}=1$ string landscape, the total gauge rank can be either 4, 12, or 20; this limitation can be understood as a quantum-gravitational consistency condition, by invoking Swampland arguments \cite{Montero:2020icj}.
Different from the rank counting in that work, which organizes the theories into having rank 2, 10, or 18, we include the contributions of the ${\cal N}=1$ gravity multiplet which always contains two graviphotons, because the associated $U(1)$ factors are generally involved in the overall gauge group topology.
} 20 in their F-theory realization \cite{Vafa:1996xn}, where the relevant geometric features \cite{Aspinwall:1998xj,Mayrhofer:2014opa,Cvetic:2017epq} have been classified \cite{shimada_k3}.

To lend further credence, but, more importantly, to collect additional ``data'' to eventually sharpen these arguments,\footnote{The arguments of \cite{Cvetic:2020kuw,Montero:2020icj} provide necessary, but not sufficient criteria for a non-trivial global gauge group structure, see \cite{Cvetic:2020kuw} for a detailed discussion.
} it would be desirable to also study other branches of the 8d moduli space.
Unfortunately, there does not exist a classification of gauge group topologies in rank 12 or rank 4 theories as comprehensive as in the case of rank 20 theories \cite{shimada_k3,Font:2020rsk}.
With this motivation in mind, the purpose of this work is to provide the general framework to determine the gauge group topology in 8d ${\cal N}=1$ string models, with a focus on rank 12 theories.

Rank 12 theories arise as $S^1$-reductions of the CHL string \cite{Chaudhuri:1995fk,Chaudhuri:1995bf}.
The physical states, which are characterized by the winding numbers and momenta of the CHL string, live in an even lattice $\Lambda_M$ of rank 12, the so-called Mikhailov lattice \cite{Mikhailov:1998si}.
Then, any non-Abelian gauge \emph{algebra} $\fkg$ that can arise in an 8d CHL vacuum must have a root lattice $\Lambda_\text{r}^\fkg$ that embeds in a specific way into $\Lambda_M$.
Such lattice embeddings can be classified \cite{Font:2021uyw} in an analogous fashion as for rank 20 theories based on their heterotic realization \cite{Font:2020rsk}, where the corresponding string momentum lattice is the rank 20 Narain lattice $\Lambda_N$ \cite{Narain:1985jj,Narain:1986am}.

On the other hand, as we will elaborate in Section \ref{sec:2}, the information about the global structure of the gauge \emph{group} $G = \widetilde{G} / {\cal Z}$, with $\widetilde{G}$ the simply-connected group with algebra $\fkg$, is encoded in the lattice \emph{dual} to the string momentum lattice $\Lambda_S$ with $\Lambda_S = \Lambda_N$ or $\Lambda_M$.
Roughly speaking, the definition of the dual lattice $\Lambda_S^* \subset \Lambda_S \otimes \mathbb{R}$ as having integer pairing with all vectors in $\Lambda_S$ can be regarded as a constraint on the representations of the physical states in $\Lambda_S$.
More precisely, the fundamental group,
\begin{align}
    {\cal Z} = \pi_1(G) = \Lambda_\text{cc}^G / \Lambda_\text{cr}^\fkg \, ,
\end{align}
depends on the \emph{cocharacter} lattice $\Lambda_\text{cc}^G$, which is a sublattice of the \emph{coweight} lattice $\Lambda_\text{cw}^\fkg = \Lambda_\text{r}^*$.
This is the dual of the character lattice $\Lambda_\text{c}^G$, which corresponds to the charge lattice occupied by physical states,\footnote{Here, we adapt the notation from \cite{Gukov:2006jk}.
It is also common (see, e.g. \cite{bump2004lie,Hall2015}) to refer to $\Lambda_\text{c}^G$ ($\Lambda_\text{cc}^G$) as the (co-)weight lattice \emph{of the group} $G$.
}
which clearly is the momentum lattice $\Lambda_S$.
From this perspective, the self-duality of the Narain lattice  (imposed by modularity of the heterotic worldsheet), together with the fact that rank 20 theories only have ADE-algebras (whose (co-)root lattices $\Lambda_\text{r}^\fkg = \Lambda_\text{cr}^\fkg$ agree), appear as a coincidence that makes it straightforward to compute the fundamental group ${\cal Z} = \pi_1(G)$ as (the torsional piece\footnote{The free part corresponds to $U(1)$ symmetries, which in fact can also have a non-trivial global structure with the non-Abelian group; we will elaborate on this in detail below.} of) $\Lambda_N / \Lambda_\text{r}^\fkg$, as done in \cite{Font:2020rsk}.
This is confirmed via duality by F-theory geometries \cite{shimada_k3}, where the corresponding data are encoded in the Mordell--Weil group \cite{Aspinwall:1998xj,Mayrhofer:2014opa,Cvetic:2017epq}.
In the rank 12 case, this quotient is no longer the correct object to compute, due to $\Lambda_M \neq \Lambda_M^*$, as well as the appearance of non-simply laced $\mathfrak{sp}$ algebras with $\Lambda_\text{r}^\mathfrak{sp} \neq \Lambda_\text{cr}^\mathfrak{sp}$.
Instead, as we shall demonstrate explicitly in Section \ref{sec:2}, the correct prescription for $\pi_1(G)$ of CHL vacua is captured by the ``mismatch'' between $\Lambda_\text{cr}^\fkg$ and $\Lambda_M^*$.

Moreover, our approach naturally computes the global gauge group structure including the $U(1)$ gauge factors.
That is, given the embedding data $\Lambda_\text{r}^\fkg \subset \Lambda_S$ of the non-Abelian root lattice into the momentum lattice, we can determine the entire gauge group topology, which takes the generic form
\begin{align}
    \frac{[\widetilde{G}/{\cal Z}] \times U(1)^{r_F}}{{\cal Z}'} \, ,
\end{align}
with $r_F = \text{rank}(\Lambda_S) - \text{rank}(\fkg)$.
As we will explain, the quotient ${\cal Z}'$, which may be interpreted as a constraint on the $U(1)$ charges of states in certain representations of $\fkg$, arises due to lattice generators of $\Lambda_S^*$ that are not in the plane containing $\Lambda_\text{r}^\fkg$.
For rank 20 theories, our approach is equivalent to methods based on string junctions that describe the dual F-theory model \cite{Guralnik:2001jh,junctions_tba}, and we will demonstrate its efficacy also in a concrete CHL model below.

An important consequence, which we prove in Section \ref{subsec:1-form_anomalies}, is that the non-Abelian gauge group topology $\widetilde{G}/{\cal Z}$ is consistent with a gauged 1-form ${\cal Z}$ symmetry \cite{Gaiotto:2014kfa}, in both heterotic and CHL vacua.
That is, there is no mixed anomaly that would obstruct such a gauging, consistent with the findings in \cite{Cvetic:2020kuw}.
We verify this explicitly by computing ${\cal Z} = \pi_1(G)$ for all maximally enhanced CHL models (i.e., those with rank$(G)=10$), which is presented in Appendix \ref{app:big_table}. 
We also find a consistent cross-check for two of these models, which are subject to constraints posed in \cite{Montero:2020icj}.
To facilitate the computation of ${\cal Z}$, we show, in Section \ref{sec:3}, that for any CHL model, specified by an embedding $\Lambda_\text{r}^\fkg \subset \Lambda_M$, the corresponding gauge group topology can be directly inferred from that of a ``parent'' rank 20 heterotic model with $G_\text{het} = \widetilde{G}_\text{het} / {\cal Z}_\text{het}$.\footnote{Via string dualities, the CHL model corresponds to IIB with an O7$_+$ plane, or, equivalently, F-theory on a K3-surface with a (partly) ``frozen'' singularity \cite{Witten:1997bs,Tachikawa:2015wka,Bhardwaj:2018jgp}.
The same K3, when interpreted without the frozen singularity, defines a rank 20 F-theory model that is dual to the ``parent'' heterotic model.
}
This then allows for an easy extraction of ${\cal Z}$ via ${\cal Z}_\text{het}$, the latter of which can be obtained from the heterotic classification \cite{Font:2020rsk,junctions_tba}.
We conclude in Section \ref{sec:conclusions} with some outlook to related topics.

\section{Gauge Groups from Momentum Lattices}
\label{sec:2}

We begin this section by reviewing the group-theoretic definition of the global gauge group structure in terms of the various lattices.
We then discuss how these structures emerge in root lattice embeddings into the momentum lattice $\Lambda_S$ of string states.
We will highlight the key differences between rank 20 heterotic theories with $\Lambda_S = \Lambda_N$ the Narain lattice, and rank 12 CHL theories with $\Lambda_S = \Lambda_M$ the Mikhailov lattice.

\subsection{Lattices and Gauge Group Topology}

Any non-Abelian gauge algebra $\mathfrak{g}$ of rank $r$ is specified by a root system $\Phi_\mathfrak{g}$, which is a finite subset of a Euclidean vector space $E \cong \mathbb{R}^r$ satisfying certain axioms (see, e.g., \cite{bump2004lie,Hall2015} for a broader introduction; we follow the conventions of \cite{Gukov:2006jk}). 
Relevant to us in the following will be that the root lattice $\Lambda^\fkg_\text{r} \supset \Phi_\fkg$ --- spanned by integer linear combinations of simple roots $\boldsymbol{\mu} \in \Phi_\fkg$ --- is a rank $r$ lattice inside $E$.
The space $E$ comes equipped with a bilinear pairing $( \cdot , \cdot ) : E \times E \rightarrow \mathbb{R}$ which induces a pairing on $\Lambda^\fkg_\text{r}$.
The normalization is such that $(\boldsymbol{\nu}, \boldsymbol\nu)= 2$ for $\boldsymbol\nu \in \Phi$ a short root, and $(\boldsymbol\nu, \boldsymbol\nu)= 4$ for the long root of $\mathfrak{sp}(n)$.
The axioms also assert that $2(\boldsymbol\nu_1, \boldsymbol\nu_2)/(\boldsymbol\nu_2, \boldsymbol\nu_2) \in \bbZ$ for any two roots $\boldsymbol\nu_1, \boldsymbol\nu_2 \in \Phi_\fkg$, ensuring that the coroots,
\begin{align}\label{eq:coroots_def}
    \Phi_\fkg^\vee =\left\{ \left. \boldsymbol\nu^\vee := \frac{2\boldsymbol\nu}{(\boldsymbol\nu, \boldsymbol\nu)} \, \right| \, \boldsymbol\nu \in \Phi_\fkg \right\} \subset E \, ,
\end{align}
and their integer span $\Lambda^\fkg_\text{cr}$, the coroot lattice, have integer pairings with roots.
For $\fkg$ an ADE algebra, we have $\Lambda^\fkg_\text{r} = \Lambda^\fkg_\text{cr}$, because all ADE roots have length squared 2.
One then defines the weight and coweight lattices, $\Lambda^\fkg_\text{w}$ and $\Lambda^\fkg_\text{cw}$, as their respective dual lattices:\footnote{Given a lattice $\Lambda$ with pairing $(\cdot, \cdot)$, the dual lattice is defined to be $\Lambda^* = \{\overline{\bf a} \in \Lambda \otimes \mathbb{R} \, | \, (\overline{\bf a}, {\bf v}) \in \bbZ \, \text{ for all } \, {\bf v} \in \Lambda \}$.
$\Lambda^*$ has the same rank as $\Lambda$.
}
\begin{align}\label{eq:weight_coweight_lattice_def}
    \begin{split}
        \Lambda^\fkg_\text{w} & := (\Lambda^\fkg_\text{cr})^* = \{ {\bf w} \in E \, | \, ({\bf w}, \boldsymbol\alpha^\vee) \in \bbZ \, \text{ for all } \, \boldsymbol\alpha^\vee \in \Lambda^\fkg_\text{cr} \} \supset \Lambda^\fkg_\text{r} \, ,\\
        \Lambda^\fkg_\text{cw} & := (\Lambda^\fkg_\text{r})^* = \{ \overline{\bf w} \in E \, | \, (\overline{\bf w}, \boldsymbol\alpha) \in \bbZ \, \text{ for all }\, \boldsymbol\alpha \in \Lambda^\fkg_\text{r} \} \supset \Lambda^\fkg_\text{cr} \, .
    \end{split}
\end{align}
Note that all these lattices are of rank $r$, i.e., they span $E$ over $\mathbb{R}$.
If $\fkg = \oplus_j \fkg_j$ is a sum of simple factors, there is an orthogonal decomposition $E = \oplus_j E_j$, where $E_j$ are spanned by the roots $\Phi_{\fkg_j}$ and their associated lattices of the corresponding simple factor $\fkg_j$.

So far, all data are defined by the gauge algebra $\mathfrak{g}$ with roots $\Phi_\fkg$.
The actual gauge group $G$ is specified by a third pair of lattices, the character lattice $\Lambda^G_\text{c}$ and the cocharacter lattice $\Lambda^G_\text{cc}$, which are intermediate lattices,
\begin{align}\label{eq:character_cochar_lattice_def}
    \begin{split}
        \Lambda^\fkg_\text{r} \subset\, & \Lambda^G_\text{c} \subset \Lambda^\fkg_\text{w} \, , \\
        \Lambda^\fkg_\text{cr} \subset\, & \Lambda^G_\text{cc} \subset \Lambda^\fkg_\text{cw} \, ,
    \end{split}
\end{align}
that are dual to each other, $(\Lambda^G_\text{c})^* = \Lambda^G_\text{cc}$, with respect to the pairing $( \cdot, \cdot)$.
A gauge theory with group $G$ can only have dynamical states whose weight vectors lie in $\Lambda_\text{c}^G$, which is also often called the weight lattice \emph{of the group} $G$.\footnote{One can show, see, e.g., \cite{bump2004lie}, that $\Lambda^G_\text{c}$ is isomorphic to character group Hom$(T, \mathbb{C}^\times)$ of the maximal torus $T \subset G$ of the \emph{group}.
}
In terms of the (co-)character lattices, the center and the fundamental group of $G$ are:
\begin{align}\label{eq:group_topology_via_lattices_general}
\begin{split}
    Z(G) & = \Lambda^\fkg_\text{cw} / \Lambda^G_\text{cc} \cong \Lambda^G_\text{c} / \Lambda^\fkg_\text{r}\, , \\
    \pi_1(G) & = \Lambda^G_\text{cc} / \Lambda^\fkg_\text{cr} \cong \Lambda^\fkg_\text{w} / \Lambda^G_\text{c}\, .
\end{split}
\end{align}

If $G = \widetilde{G}$ is the simply-connected group with algebra $\fkg$, then $\Lambda^{\widetilde{G}}_\text{c} = \Lambda^\fkg_\text{w}$ and $\Lambda^{\widetilde{G}}_\text{cc} = \Lambda^\fkg_\text{cr}$.
Elements $c$ in the center $Z(\widetilde{G}) = \Lambda^\fkg_\text{cw} / \Lambda^\fkg_\text{cr}$, represented by a coweight $\overline{\bf v}_c \in \Lambda^\fkg_\text{cw}$, act on a weight by a phase $\exp(2\pi i c({\bf w}))$, where the fractional number
\begin{align}\label{eq:center_charge_weight}
    c({\bf w}) = ({\bf w}, \overline{\bf v}_c) \equiv ({\bf w}, \overline{\bf v}_c + \boldsymbol{\alpha}^\vee) \mod \bbZ \quad \text{ for any } \quad \boldsymbol{\alpha}^\vee \in \Lambda_\text{cr}^\fkg \, ,
\end{align}
can be interpreted as the charge of ${\bf w}$ under the center element $c$ represented by $\overline{\bf v}_c \mod \Lambda^\fkg_\text{cr}$.
Note that this center charge is invariant for all weights of an irreducible representation ${\bf R}$ of $\fkg$, because $c({\bf w} + \boldsymbol{\alpha}) = ({\bf w} + \boldsymbol{\alpha}, \overline{\bf v}_c) = ({\bf w}, \overline{\bf v}_c) \mod \bbZ$ for roots $\boldsymbol{\alpha} \in \Lambda_\text{r}^\fkg$.

Since $\Lambda_\text{c}^G \subset \Lambda_\text{w}^\fkg \equiv \Lambda_\text{c}^{\widetilde{G}}$, we can regard a character ${\bf w} \in \Lambda_\text{c}^G$ of a non-simply connected group $G$ as weights of $\widetilde{G}$.
Then we see that they are acted on trivially by $\pi_1(G) = \Lambda_\text{cc}^G / \Lambda_\text{cr}^\fkg \subset \Lambda_\text{cw}^\fkg / \Lambda_\text{cr}^\fkg = Z(\widetilde{G})$, because they have center charges $c({\bf w}) = ({\bf w}, \overline{\bf v}_c) = 0 \mod \bbZ$ for $\overline{\bf v}_c \in \Lambda_\text{cc}^G$.
Hence, we can also view the ``non-trivial global structure'' $G = \widetilde{G} / \pi_1(G)$ of a gauge group as imposed by requiring a subgroup ${\cal Z} \equiv \pi_1(G) \subset Z(\widetilde{G})$ to act trivially on all dynamical representations.

\subsection{Gauge Group Topology from Lattice Embeddings}\label{subsec:gauge_group_from_lattice}

Compactifications of the heterotic or CHL string to 8d are characterized by a lattice $\Lambda_S$ with a symmetric non-degenerate bilinear pairing $\langle \cdot, \cdot \rangle_S: \Lambda_S \times \Lambda_S \rightarrow \bbZ$ of signature $(2,R)$.
For the heterotic string, $\Lambda_S$ is the rank 20 Narain lattice $\Lambda_N$ with $R=18$ \cite{Narain:1985jj,Narain:1986am}.
For the CHL string, $\Lambda_S$ is the rank 12 Mikhailov lattice $\Lambda_M$ with $R=10$ \cite{Mikhailov:1998si}.
In either case, we can linearly extend $\Lambda_S$ to vector space $V$ with a symmetric non-degenerate bilinear pairing $\langle \cdot , \cdot \rangle$:
\begin{align}
    V_S := \Lambda_S \otimes \mathbb{R} \, , \quad \langle \lambda_1 {\bf v}_1, \lambda_2 {\bf v}_2\rangle = \lambda_1 \lambda_2 \langle {\bf v}_1, {\bf v}_2\rangle_S \quad \text{for} \quad {\bf v}_1, {\bf v}_2 \in \Lambda_S, \, \, \lambda_1, \lambda_2 \in \mathbb{R} .
\end{align}
Since $\Lambda_S \subset V_S$, we will identify the lattice pairing $\langle \cdot , \cdot \rangle_S$ with the vector space pairing $\langle \cdot , \cdot \rangle$ in the following.
Then there is a dual lattice $\Lambda_S^* \subset V_S$ defined with respect to $\langle \cdot , \cdot \rangle$.
The Narain lattice is self-dual, $\Lambda_N^* = \Lambda_N$, but for the Mikhailov lattice, $\Lambda_M^* \neq \Lambda_M$.

By tuning the compactification moduli, the gauge symmetry of the effective theory in 8d can change.
Roughly speaking, this tuning amounts to setting the masses of certain states to 0, which can furnish the W-bosons of non-Abelian gauge symmetries.
The question of which non-Abelian gauge algebras $\fkg$ are realizable in this way can be answered by cataloging all embeddings of the root lattices $\Lambda_\text{r}^\fkg$ into $\Lambda_S$, whose roots $\Phi_\fkg$ satisfy the worldsheet conditions which guarantee their masslessness.
This process has been recently carried out in detail \cite{Fraiman:2018ebo,Font:2020rsk,Font:2021uyw}, which in particular resulted in the full list of realizable gauge algebras with maximal rank (i.e., rank$(\fkg) = 18$ for $\Lambda_\text{r}^\fkg \hookrightarrow \Lambda_N$, and rank$(\fkg)=10$ for $\Lambda_\text{r}^\fkg \hookrightarrow \Lambda_M$).

The purpose of this work is not to reiterate the necessary and sufficient criteria to find such embeddings, but to focus on the extraction of the \emph{global form} of the gauge group from the embedding data.
To this end, our working assumption will be that any root lattice embedding $\Lambda_\text{r}^\fkg \stackrel{\imath}{\hookrightarrow} \Lambda_S$ we consider in the following satisfies these criteria, which guarantees that the corresponding 8d compactification (be it heterotic or CHL) has a non-Abelian symmetry algebra $\fkg$. 
From this starting point, let us now distill the properties pertaining to the gauge group topology.

At the level of vector spaces we introduced above, an embedding $\Lambda_\text{r}^\fkg \stackrel{\imath}{\hookrightarrow} \Lambda_S$ extends to an injective homomorphism $\imath: E \hookrightarrow V_S$, with $E = \Lambda_\text{r}^{\mathfrak{g}} \otimes \mathbb{R}$, such that
\begin{enumerate}
    \item \label{crit1} $\langle (\imath({\bf v}), \imath({\bf w}) \rangle = ({\bf v}, {\bf w})$ for any ${\bf v}, {\bf w} \in E$;
    \item \label{crit2} $\imath(\Lambda^\fkg_\text{r})$ is a sublattice of $\Lambda_S \subset V_S$;
    \item \label{crit3} $\imath(\Lambda^\fkg_\text{cr})$ is a sublattice of $\Lambda_S^* \subset V_S$.
\end{enumerate}
The first and second points are just a careful restatement of ``$\Lambda_\text{r}^\fkg \stackrel{\imath}{\hookrightarrow} \Lambda_S$ is a lattice embedding''.
For heterotic vacua, the third point is equivalent to the second, since $\Lambda_N = \Lambda_N^*$, and $\Lambda_\text{r}^\fkg = \Lambda_\text{cr}^\fkg$ for an ADE algebra $\fkg$.
For the CHL string this is a non-trivial criterion, which however is satisfied in valid embeddings \cite{Mikhailov:1998si}, as we will discuss below.
From criterion \ref{crit1}, it is straightforward to show that $\imath(\Lambda^*) = \imath(\Lambda)^*$ for any lattice $\Lambda \subset E$.
Then, the second and third conditions imply $\imath(\Lambda_\text{cw}^\fkg) = \imath((\Lambda_\text{r}^\fkg)^*) \supset \Lambda_S^* \cap \imath(E)$, and $\imath(\Lambda^\fkg_\text{w}) = \imath((\Lambda_\text{cr}^\fkg)^*) \supset \Lambda_S \cap \imath(E)$.

Given such an embedding $\imath: E \hookrightarrow V_S$, we naturally have an orthogonal decomposition 
\begin{align}\label{eq:decomp_V_into_E+F}
    V_S = \imath(E) \oplus F \, , \quad \text{where} \quad F = \{{\bf v} \in V \, | \, \langle {\bf v}, \imath({\bf w}) \rangle = 0 \, \text{ for all } \, {\bf w} \in E \} \, ,
\end{align}
because the restriction of $\langle \cdot , \cdot \rangle$ to $\imath(E)$ is the pairing $(\cdot, \cdot)$ which is non-degenerate.
For later convenience, we define the projections 
\begin{align}
\pi_F: \imath(E) \oplus F &\rightarrow F , \\
\pi_E: \imath(E) \oplus F &\rightarrow \imath(E).
\end{align} This decomposition determines the number of independent $\mathfrak{u}(1)$ gauge factors to be $\dim_\mathbb{R} (F) \equiv r_F= 2+R - \text{rank}(\mathfrak{g})$.

The lattice points of $\Lambda_S \subset V_S$ define physical states, and lattice points of $\Lambda_S^*$ impose constraints on the $\mathfrak{g}$-representations and $\mathfrak{u}(1)$ charges of these states, because they have to pair integrally with points in $\Lambda_S$.
These constraints can be interpreted as a non-trivial global structure of the gauge group of the form
\begin{align}\label{eq:global_gauge_group_general}
    \frac{[\widetilde{G}/{\cal Z}] \times U(1)^{r_F}}{{\cal Z}'} \, ,
\end{align}
where $\widetilde{G}$ is the simply-conncted version of the non-Abelian group with algebra $\mathfrak{g}$, ${\cal Z} \subset Z(\widetilde{G})$ a subgroup of the center, and ${\cal Z}'$ embeds into both $Z(\widetilde{G})$ and $U(1)^{r_F}$.

Let us first understand the ``purely non-Abelian'' constraints, i.e., those that specify the non-Abelian group $G = \widetilde{G} / {\cal Z}$.
These are restrictions on the physically realized weights that form the character lattice $\Lambda^G_\text{c} \subset \Lambda_\text{w}^\fkg \subset E$.
In the string realization, any physical state corresponds to a lattice point ${\bf s} \in \Lambda_S$, which can be decomposed orthogonally as ${\bf s} = {\bf s}_E + {\bf s}_F \in \imath(E) \oplus F$.
The weight ${\bf w} \in \Lambda_\text{w} \subset E$ of such a state ${\bf s}$ under the non-Abelian part $G = \widetilde{G} / {\cal Z}$ is then the orthogonal projection of $\bf s$ onto $\imath(E)$, i.e., ${\bf s}_E  = \pi_E ({\bf s})$.\footnote{
More precisely, we identify ${\bf s}_E = \imath({\bf w})$.
Recalling that any weight is specified by its Dynkin labels ${\bf w}_i = ({\bf w}, \boldsymbol\mu_i^\vee)$, where $\boldsymbol\mu_i^\vee \in \Lambda_\text{cr} \subset E$ are the simple coroots, we have $\langle {\bf s}, \imath(\boldsymbol\mu_i^\vee) \rangle = \langle {\bf s}_E, \imath(\boldsymbol\mu_i^\vee) \rangle = ({\bf w}, \boldsymbol\mu_i^\vee)$.}

In other words, \textit{the character lattice of $G$ is the orthogonal projection of $\Lambda_S$ onto $\imath(E)$}: 
\begin{align}\label{eq:char_lat_non-ab}
    \Lambda^G_\text{c} \cong \pi_E (\Lambda_S) \subset \imath(E) \, .
\end{align}
The vectors ${\bf s} \in \Lambda_S$ are subject to the constraint that they pair integrally with all points in $\Lambda_S^*$.
Consider in particular a constraint associated with a point ${\bf c} \in \Lambda_S^* \cap \imath(E) \subset \imath(\Lambda_\text{cw})$, and let $\overline{\bf v} \in \Lambda_\text{cw}$ be such that $\imath(\overline{\bf v}) = {\bf c}$.
By orthogonality, we have 
\begin{align}
    \langle {\bf s}, {\bf c} \rangle = \langle\pi_E({\bf s}) , {\bf c}\rangle = \langle \imath({\bf w}), \imath(\overline{\bf v})\rangle = ({\bf w}, \overline{\bf v}) \in \bbZ \, .
\end{align}
This shows that $\Lambda_S^* \cap \imath(E)$ can be identified with the cocharacter lattice $\Lambda_\text{cc}^G$ of $G$.
So, from \eqref{eq:group_topology_via_lattices_general}, the non-Abelian gauge \emph{group} $G$ satisfies
\begin{align}\label{eq:non_ab_group_from_string_lat}
    \begin{split}
        Z(G) &= \frac{\Lambda^G_\text{c}}{\Lambda^\fkg_\text{r}} = \frac{\pi_E(\Lambda_S)}{\imath(\Lambda^\fkg_\text{r})} \, , \qquad \pi_1(G) = \frac{\Lambda^G_\text{cc}}{\Lambda^\fkg_\text{cr}} = \frac{\Lambda_S^* \cap \imath(E)}{\imath(\Lambda^\fkg_\text{cr})} \, . 
    \end{split}
\end{align}

Equivalently to \eqref{eq:char_lat_non-ab}, the projection of the lattice $\Lambda_S$ of physical states onto $F$ gives the ``characters'' of the $U(1)$s, i.e., the possible $U(1)$ charges.
Just as how the non-Abelian weight ${\bf w}({\bf s})$ of a state is specified by the Dynkin labels ${\bf w}_i = ({\bf w}, \boldsymbol\mu_i^\vee) = \langle {\bf s}_E, \imath(\boldsymbol\mu_i^\vee) \rangle$, where the simple coroots $\boldsymbol\mu_i^\vee$ span $E$ (over $\mathbb{R}$), the $U(1)$ charges are defined by the pairing with basis vectors of $F$.
To fix the normalization of the $U(1)$s, we use a lattice basis $\boldsymbol\xi_{\ell}$, $\ell=1,..., r_F$, for $\Lambda^*_S \cap F$, i.e., the orthogonal complement of $\imath(\Lambda^\fkg_\text{cr})$ inside $\Lambda_S^*$ (we will see momentarily that $\Lambda^*_S \cap F \neq \emptyset$):
\begin{align}
    q_\ell({\bf s}) := \langle {\bf s}, \boldsymbol\xi_\ell \rangle \, .
\end{align}
In this normalization, states ${\bf s} \in \Lambda_S$ that are singlets under the non-Abelian gauge algebra $\fkg$, i.e., $\pi_E({\bf s}) = 0 \Leftrightarrow {\bf s} \in \Lambda_S \cap F$, clearly have integer $U(1)$ charges $q_\ell$. 
The lattice points of $\Lambda_S^*$ that are not inside $\imath(E)$ now constrain the $U(1)$-charges $q_i({\bf s})$ and the non-Abelian weights ${\bf w}({\bf s})$ of a physical state corresponding to ${\bf s} \in \Lambda_S$.
To see this, we orthogonally decompose $\Lambda_S^* \ni {\bf c} = {\bf c}_E + {\bf c}_F$.
Note that, in general, neither ${\bf c}_E$ nor ${\bf c}_F$ are lattice points of $\Lambda_S^*$!
But, because for a root $\imath(\boldsymbol\alpha) \in \imath(\Lambda^\fkg_\text{r}) \subset \Lambda_S \cap \imath(E)$, we have $\bbZ \ni \langle {\bf c}, \imath(\boldsymbol\mu) \rangle = \langle {\bf c}_E, \imath(\boldsymbol\mu) \rangle$, this guarantees that ${\bf c}_E = \imath(\overline{\bf v}) \in \imath(\Lambda^\fkg_\text{cw})$ for some coweight $\overline{\bf v}$ of $\mathfrak{g}$.

Then, since $\Lambda^\fkg_\text{cr} \subset \Lambda^\fkg_\text{cw}$ are lattices of the same rank, we know that for any $\overline{\bf v} \in \Lambda^\fkg_\text{cw}$ there is a smallest positive integer $k$ such that $\imath(k \overline{\bf v}) = k {\bf c}_E \in \imath(\Lambda^\fkg_\text{cr}) \subset \Lambda_S^*$, so $k {\bf c}_F = k {\bf c} - k{\bf c}_E \in \Lambda_S^* \cap F$ is an \textit{integer} linear combination of $\boldsymbol\xi_{\ell}$. 
This means that $\langle {\bf c}_F, {\bf s} \rangle = \sum_\ell \lambda_\ell q_\ell({\bf s})$ is a $k$-fractional linear combination of the $U(1)$ charges $q_\ell({\bf s})$ of ${\bf s}$.
Therefore, the vector ${\bf c} \in \Lambda_S^*$ of the dual lattice imposes that 
\begin{align}
    \sum_\ell \lambda_\ell q_\ell({\bf s}) + ({\bf w}({\bf s}), \overline{\bf v}) \in \bbZ \, .
    \label{eq:cocharacter_constraint}
\end{align}
Moreover, from the above considerations it is clear that $k\lambda_\ell \in \bbZ$ and $k({\bf w}({\bf s}), \overline{\bf v})=({\bf w}({\bf s}), k\overline{\bf v}) \in \bbZ$.
Hence, the constraint is a $\bbZ_k$ constraint, in that it becomes trivial when it is multiplied by $k$.
It can be interpreted as identifying a $\bbZ_k \subset Z(\widetilde{G})$ with a subgroup of $U(1)^{r_F}$, i.e., it defines a $\bbZ_k$ subgroup of ${\cal Z}'$ in \eqref{eq:global_gauge_group_general}.
Just by counting dimension of $\Lambda_S^* / (\Lambda_S^* \cap \imath(E)$), there are at most $r_F$ linearly independent such constraints that are also independent of the ``non-Abelian constraints'' in ${\cal Z}$, i.e., ${\cal Z}' \cong \prod_{\ell=1}^{r_F} \bbZ_{k_\ell}$.
Then, analogously to \eqref{eq:non_ab_group_from_string_lat}, we have
\begin{align}\label{eq:ab_group_from_string_lat}
    {\cal Z}' \cong \frac{\Lambda_\text{cc}'}{\imath(\Lambda^\fkg_\text{cr})} \quad \text{with} \quad \Lambda_\text{cc}' := \pi_E(\Lambda_S^*) \subset \imath(\Lambda^\fkg_\text{cw}) \, .
\end{align}

In general, $\fkg = \oplus \fkg_i$ will be a sum of simple algebras, with a orthogonal decomposition of the lattice $\Lambda_\text{cw}^\fkg = \oplus_i \Lambda_\text{cw}^{\fkg_i}$.
Then, any lattice vector ${\bf c} \in \Lambda_\text{cc}'$ or ${\bf c} \in \Lambda_\text{cc}^G$ has a unique decomposition ${\bf c} = \sum_i \imath(\overline{\bf w}_i)$, where $\overline{\bf w}_i \in \Lambda_\text{cw}^{\fkg_i}$ defines an equivalence class $[\overline{\bf w}_i] \equiv k_i \in \Lambda_\text{cw}^{\fkg_i} / \Lambda_\text{cr}^{\fkg_i} = Z(\widetilde{G}_i)$.
Then, the equivalence class of ${\bf c}$ in $\Lambda_\text{cc}' / \imath(\Lambda_\text{cw}^\fkg)$ (or $\Lambda_\text{cc}^G / \imath(\Lambda_\text{cw}^\fkg)$) $\subset \imath(\Lambda_\text{cw}^\fkg) / \imath(\Lambda_\text{cr}^\fkg) \cong Z(\widetilde{G})$ can be represented by the tuple $(k_i) \in \prod_i Z(\widetilde{G}_i) = Z(\widetilde{G})$.

In the following, we will exemplify the above structures in 8d heterotic and CHL vacua.
To ease the notation, we will from now on drop the explicit embedding map $\imath$, and regard all occurring lattices and subspaces as embedded into $V_S := \Lambda_S \otimes \mathbb{R} = \Lambda_S^* \otimes \mathbb{R}$, with all pairings inherited from $\langle \cdot, \cdot \rangle$ on $V_S$.

\subsubsection*{Global gauge group structure of 8d heterotic vacua}

For 8d heterotic vacua with gauge rank 20, the topology of the non-Abelian gauge symmetry $G = \widetilde{G} / {\cal Z}$ has been recently studied through lattice embeddings in $\Lambda_N$ in \cite{Font:2020rsk}.
There, the crucial data were an overlattice $M$ of the root lattice $\Lambda_\text{r}^\fkg$, whose length-squared 2 lattice points coincide with $\Lambda_\text{r}^\fkg$, that embeds primitively inside $\Lambda_N$.
Then, the identification ${\cal Z} = \pi_1(G) = M / \Lambda_\text{r}^\fkg$ was cross-checked with the classification of Mordell--Weil torsion of elliptic K3-surfaces in \cite{shimada_k3}, which is known to provide an equivalent characterization of the non-Abelian gauge group topology of 8d heterotic vacua via F-theory \cite{Aspinwall:1998xj}.

Comparing with the general formula \eqref{eq:group_topology_via_lattices_general} for $\pi_1(G)$, this identification seems to be at odds at first, since it is the \textit{coroot lattice} $\Lambda^\fkg_\text{cr}$ rather than root lattice $\Lambda^\fkg_\text{r}$ that appears in the quotients characterizing the fundamental group.
Of course, this is remedied by the fact that, in 8d heterotic vacua, only ADE algebras $\fkg$ can be realized, which have $\Lambda^\fkg_\text{r} = \Lambda^\fkg_\text{cr}$.
Then, to be consistent with \eqref{eq:group_topology_via_lattices_general}, the overlattice $M$ should be identified with the cocharacter lattice $\Lambda_\text{cc}^G$.
Indeed, because $M$ contains $\Lambda^\fkg_\text{r} = \Lambda^\fkg_\text{cr}$, the requirement that it embeds primitively into $\Lambda_N$ means that it contains all points of $\Lambda_N \cap E$.
Furthermore, as $M/\Lambda_\text{r}^\fkg$ is of finite order, $M$ has the same rank as $\Lambda_\text{r}^\fkg$, which is the same as the dimension of $E$, so it cannot contain more points than $\Lambda_N \cap E$.
Therefore, we indeed find $M = \Lambda_N \cap E = \Lambda_N^* \cap E$ to be the cocharacter lattice as in \eqref{eq:non_ab_group_from_string_lat}. 

Our proposal can further determine the non-trivial constraints ${\cal Z}'$ between the non-Abelian group and the $U(1)$s of the heterotic compactification.
Note that, through duality to F-theory, there is an independent method to determine this structure via string junctions \cite{Guralnik:2001jh}.
While we leave a full proof of the equivalence between these two methods to an upcoming work \cite{junctions_tba}, we remark here that we indeed find identical results for 8d heterotic string vacua.
We will present, for completeness, an example of a heterotic model with $\fkg = \mathfrak{su}(2)^2 \oplus \mathfrak{su}(4)^2 \oplus \mathfrak{so}(20)$ in Appendix \ref{app:heterotic_example}, where we show that the global gauge group is
\begin{align}\label{eq:heterotic_example_gauge_group}
    \frac{[SU(2)^2 \times SU(4)^2 \times Spin(20)]/[\bbZ_2 \times \bbZ_2] \times U(1)^2}{\bbZ_4 \times \bbZ_4} \, .
\end{align}

\subsection{Gauge Group Topology of 8d CHL Vacua}
\label{subsec:CHL_example}

Our main focus is to derive the gauge group topology of 8d CHL vacua.
The important difference from heterotic vacua is the fact that the string lattice $\Lambda_S$ is no longer self-dual in this case.
As found in \cite{Mikhailov:1998si}, the rank 12 momentum lattice is the Mikhailov lattice
\begin{align}
    \Lambda_M = U(2) \oplus U \oplus \text{E}_8 \cong U \oplus U \oplus \text{D}_8 \, .
\end{align}
Here, $\text{E}_8$ (D$_8$) denotes the root lattice of the Lie group $E_8$ ($Spin(16)$).
The rank 2 lattice $U = \{ l {\bf e} + n {\bf f} \, | \, (n,l) \in \bbZ^2 \}$ is defined by the Gram matrix
\begin{align}\label{eq:gram_matrix_U_lattice}
    \begin{pmatrix} \langle {\bf e}, {\bf e}\rangle & \langle {\bf e}, {\bf f}\rangle \\ \langle {\bf f}, {\bf e}\rangle & \langle {\bf f}, {\bf f} \rangle \end{pmatrix}  = \begin{pmatrix} 0 & 1 \\ 1 & 0 \end{pmatrix} \,.
\end{align} 
In this basis, $U(2) = \{ l{\bf e} + n{\bf f} \, | \, l \in 2\bbZ, n \in \bbZ \}$ . 
The dual Mikhailov lattice is then
\begin{align}
    \Lambda_M^* = \overline{U}(\tfrac{1}{2}) \oplus U \oplus \text{E}_8 \cong U \oplus U \oplus \text{D}_8^* \, ,
\end{align}
with $\overline{U}(\tfrac{1}{2})= \{ l{\bf e} + n{\bf f} \, | \, l \in \bbZ, n \in \tfrac12 \bbZ \}$.

The criteria for embeddings of root lattices into $\Lambda_M$ have also been studied in \cite{Mikhailov:1998si}.
One key novelty, compared to heterotic vacua, is that one can realize non-simply-laced $\mathfrak{sp}(n)$ gauge algebras in 8d CHL vacua.
Importantly, one criterion of the associated root lattice embedding $\Lambda_\text{r}^{\mathfrak{sp}(n)} \stackrel{\imath}{\hookrightarrow} \Lambda_M$ is that a long root $\boldsymbol\nu_L$ (with $( \boldsymbol\nu_L, \boldsymbol\nu_L) = 4$) of $\mathfrak{sp}(n)$ must embed such that it has \emph{even} pairing with all points in $\Lambda_M$:
\begin{align}
    \langle \imath(\boldsymbol{\nu}_L) , {\bf v} \rangle \in 2\bbZ \quad \text{for all} \quad {\bf v} \in \Lambda_M \, .
\end{align}
This in turn means that the short coroots, $\boldsymbol\nu_L^\vee = 2\boldsymbol\nu_L / (\boldsymbol\nu_L , \boldsymbol\nu_L) = \frac12 \boldsymbol\nu_L$, pair integrally with $\Lambda_M$.
In particular, this means $\boldsymbol\nu_L^\vee \in \Lambda_M^*$.
Since all other roots have length 2, and thus map to themselves as coroots, the coroot lattice $\Lambda_\text{cr}^{\mathfrak{sp}(n)}$ is guaranteed to embed into $\Lambda_M^*$, which is our condition \ref{crit3} for the embedding map $\imath: E \hookrightarrow V$.
As a result, the methods outlined in Section \ref{subsec:gauge_group_from_lattice} carry through.

To highlight the difference from the process for heterotic vacua, note that, in general, the ``overlattice'' $\Lambda_M \cap E$ of the root lattice $\Lambda_\text{r}^\fkg \subset E$ neither contains all points of actual cocharacter lattice $\Lambda_\text{cc}^G = \Lambda_M^* \cap E$, nor those of the character lattice $\Lambda_\text{c}^G = \pi_E (\Lambda_M)$. 
For example, this discrepancy means that the quotient $(\Lambda_M \cap E) / \Lambda_\text{r}^\fkg$ generally gives only a \emph{subgroup} of the center $Z(G) = \pi_E (\Lambda_M) / \Lambda_\text{r}^\fkg$, and is not directly related to the fundamental group $\pi_1(G)$.

\subsubsection*{Example}

To explicitly demonstrate our approach, we will consider a CHL model with $\fkg = \mathfrak{su}(2)^2 \oplus \mathfrak{su}(4)^2 \oplus \mathfrak{sp}(2)$.
To this end, we represent $\mathbf{v}^{(\ell)} \in V_M := \Lambda_M \otimes \mathbb{R}$
\begin{align}
    \mathbf{v}^{(\ell)} = (l^{(\ell)}_1, l^{(\ell)}_2, n^{(\ell)}_1, n^{(\ell)}_2; \sigma^{(\ell)}_1, \dots, \sigma^{(\ell)}_8 ) \, ,
\end{align}
with pairing
\begin{equation}
    \langle \mathbf{v}^{(1)}, \mathbf{v}^{(2)} \rangle = l^{(1)}_1 n^{(2)}_1 + l^{(2)}_1 n^{(1)}_1 + l^{(1)}_2 n^{(2)}_2 + l^{(2)}_2 n^{(1)}_2 + \sum_{j = 1}^{8} \sigma_j^{(1)}  \sigma_j^{(2)} \, .
\end{equation}
Then, in the presentation $\Lambda_M = U \oplus U(2) \oplus \text{E}_8$ of the Mikhailov lattice, the $U$ lattice is spanned by $(l_1, 0, n_1, 0; 0, ...)$ with $l_1, n_1 \in \bbZ$, while the $U(2)$ part is spanned by $(0,l_2,0,n_2;0,...)$ with $l_2 \in 2\bbZ, n_2 \in \bbZ$ (see also \eqref{eq:gram_matrix_U_lattice}).
The $\text{E}_8$ lattice is then generated by $(0,0,0,0;\boldsymbol\sigma)$ with
\begin{align}
    \text{E}_8 \cong \left\{ \boldsymbol\sigma = (\sigma_1 , ..., \sigma_8) \in \left(\frac{1}{2}\bbZ\right)^8  \, \middle| \,  \sum_{i=1}^8 \sigma_i \in 2 \bbZ \ \ \text{and} \ \ \sigma_i - \sigma_j \in \bbZ \ \ \forall i, j \right\} \, .
\end{align}
For $\overline{\bf v} \in \Lambda_M^* = U \oplus \overline{U}(\tfrac12) \oplus \text{E}_8 \subset V_M$, the only difference for the conditions on the coefficients is that $l_2 \in \bbZ$ and $n_2 \in \tfrac12 \bbZ$.

The root lattice embedding which realizes the gauge algebra $\fkg = \mathfrak{su}(2)^2 \oplus  \mathfrak{su}(4)^2 \oplus \mathfrak{sp}(2)$ has been computed in \cite{Font:2021uyw}.
$\Lambda_\text{r}^\fkg$ is specified by the embedding of the simple roots $\boldsymbol\mu$ into $\Lambda_M$ in the above representation: 
\begin{equation}
\left[ \begin{array}{c}
\boldsymbol\mu_1 \\ \hline
\boldsymbol\mu_2 \\ \hline
\boldsymbol\mu_3 \\
\boldsymbol\mu_4 \\
\boldsymbol\mu_5 \\ \hline
\boldsymbol\mu_6 \\
\boldsymbol\mu_7 \\
\boldsymbol\mu_8 \\ \hline
\boldsymbol\mu_9 \\ 
\boldsymbol\mu_{10}
\end{array}\right]
=
\left[\begin{array}{cccc|cccccccc}
   1   & 2 &-1 &-1 & 0 & 0 & 0 & 1 & 1 & 1 & 1 &-2 \\\hline
   1   & 0 &-1 &-1 & 0 & 0 & 0 & 0 & 0 & 0 & 0 &-2 \\\hline
\ 0\ \ & 0 & 0 & 0 & 1 &-1 & 0 & 0 & 0 & 0 & 0 & 0 \\
   0   & 0 & 0 & 0 & 0 & 1 &-1 & 0 & 0 & 0 & 0 & 0 \\
   0   & 0 & 0 & 0 &-1 &-1 & 0 & 0 & 0 & 0 & 0 & 0 \\\hline
   0   & 0 & 0 & 0 & 0 & 0 & 0 & 0 & 0 & 1 &-1 & 0 \\
   0   & 0 & 0 & 0 & 0 & 0 & 0 & 0 & 1 &-1 & 0 & 0 \\
   0   & 0 & 0 & 1 & 0 & 0 & 0 &-1 &-1 & 0 & 0 & 0 \\\hline
   1   & 0 & 1 & 0 & 0 & 0 & 0 & 0 & 0 & 0 & 0 & 0 \\
   0   & 2 &-2 &-3 & 0 & 0 & 0 & 0 & 2 & 2 & 2 &-2
\end{array}\right].
\end{equation}
Here, the first two rows ${\boldsymbol\mu}_1$, ${\boldsymbol\mu}_2$ are the simple roots of $\mathfrak{su}(2)^2$, the next groups of three are the simple roots of the two $\mathfrak{su}(4)$'s, and the last two rows are simple roots of $\mathfrak{sp}(2)$, with ${\boldsymbol\mu}_{10}$ being the long root.
The corresponding coroot lattice is spanned by ${\boldsymbol\mu}^\vee_i = {\boldsymbol\mu}_i$ for $i<9$, and ${\boldsymbol\mu}_{10}^\vee = \frac12 {\boldsymbol\mu}_{10}$.
The coweight lattice is then spanned by $\overline{\bf w}_i = (C^{-1})_{ij} {\boldsymbol\mu}_j$, with $C_{ij} = \langle {\boldsymbol\mu}_i, {\boldsymbol\mu}_j \rangle$, which we re-express in terms of the coroots:
\begin{align}\label{eq:CHL_example_coweights_as_coroots}
    \begin{split}
        & \mathfrak{su}(2)^2 : \quad \overline{\bf w}_i = \tfrac12 {\boldsymbol\mu}_i^\vee \quad (i = 1, 2) \, , \\
        & \mathfrak{su}(4)^2: \quad \overline{\bf w}_{m+i} = \begin{pmatrix}
            \nicefrac34 & \nicefrac12 & \nicefrac14 \\
            \nicefrac12 & 1 & \nicefrac12 \\
            \nicefrac14 & \nicefrac12 & \nicefrac34
        \end{pmatrix}_{ij} 
            {\boldsymbol\mu}^\vee_{m+j} \quad (m=2,5) \, ,\\
        & \mathfrak{sp}(2): \quad \overline{\bf w}_9 = {\boldsymbol\mu}_9^\vee + {\boldsymbol\mu}_{10}^\vee \, , \quad \overline{\boldsymbol\mu}_{10} = \tfrac12 {\boldsymbol\mu}_9^\vee + {\boldsymbol\mu}_{10}^\vee \, .
    \end{split}
\end{align}
The orthogonal complement $F$ of $\Lambda_\text{r}^\fkg$ in $\Lambda_M \otimes \mathbb{R}$ is spanned by
\begin{align}
    \begin{split}
    \boldsymbol\xi_1 &= (-2, 0, 2, 0; 0, 0, 0, 0, 0, 0, 0, 2), \\
    \boldsymbol\xi_2 &= (2, 4, -2, -5; 0, 0, 0, 1,  3, 3, 3, -4) \\
    \langle \boldsymbol\xi_1, \boldsymbol\xi_1 \rangle &= \langle \boldsymbol\xi_2, \boldsymbol\xi_2 \rangle = -4, \ \ \langle \boldsymbol\xi_1, \boldsymbol\xi_2 \rangle = 0 \, .
    \end{split}
\end{align}
These give the generators of the two independent $U(1)$s.

As explained above, any vector of $\overline{\bf v} \in \Lambda_{M}^*$ can be written as an integer linear combination of coweight and the $U(1)$ generators:
\begin{equation}
    \overline{\bf v} = (l_1, l_2, n_1, n_1; \sigma_1, \dots, \sigma_8) = \sum_{j = 1}^{10} k_j \overline{\bf w}_{j} + m_1 \boldsymbol\xi_1 + m_2 \boldsymbol\xi_2,\ \ \ k_j \in \bbZ \, .
\end{equation}
To determine the gauge group data \eqref{eq:non_ab_group_from_string_lat} and \eqref{eq:ab_group_from_string_lat}, we then need to express the basis of $\Lambda_M^*$ in this fashion.
This is a straightforward, but rather cumbersome exercise in linear algebra.
Sparing the details, the key step is to find the generators of $\Lambda_M^*$ that are linearly independent modulo the coroots ${\boldsymbol\mu}_i^\vee$.
For $\Lambda_M^* \cap E$, where $E = \Lambda_\text{r}^\fkg \otimes \mathbb{R}$, there are two such generators:
\begin{align}
    \begin{split}
    {\bf c}_1 &= (1, 1, -1, -2; 0, 0, -1, 0, 1, 1, 1, -2)= \overline{\bf w}_2 + \overline{\bf w}_4 + \overline{\bf w}_{10}  \, ,\\
    {\bf c}_2 &= (1, 1, 0, 0; 0, 0, 0, 0, 1, 0, 0, -1) = \overline{\bf w}_1  + \overline{\bf w}_7 + (\overline{\bf w}_9 - \overline{\bf w}_{10})  \, .
    \end{split}
\end{align}
These generate $\pi_1(G) = (\Lambda_M^*\cap E) / \Lambda_\text{cr}^\fkg$ as follows.
From \eqref{eq:CHL_example_coweights_as_coroots}, we see that ${\bf c}_1$ projects onto $\overline{\bf w}_2 = \frac12 {\boldsymbol\mu}^\vee_2 \in \Lambda_\text{cw}^{\mathfrak{su}(2)}$ of the second $\mathfrak{su}(2)$ factor, which is an order two element in $\Lambda_\text{cw}^{\mathfrak{su}(2)} / \Lambda_\text{cr}^{\mathfrak{su}(2)} \cong \bbZ_2$.
Likewise, the component $\overline{\bf w}_4 = \tfrac12 {\boldsymbol\mu}_3^\vee + {\boldsymbol\mu}_4^\vee + \tfrac12 {\boldsymbol\mu}_5^\vee$ projects onto the order two element in $\Lambda_\text{cw}^{\mathfrak{su}(4)} / \Lambda_\text{cr}^{\mathfrak{su}(4)} \cong \bbZ_4$ of the first $\mathfrak{su}(4)$ factor.
Finally, the component $\overline{\bf w}_{10} = \tfrac12 {\boldsymbol\mu}_9^\vee + {\boldsymbol\mu}_{10}^\vee$ projects onto the order 2 element in $\Lambda_\text{cw}^{\mathfrak{sp}(2)}/\Lambda_\text{cr}^{\mathfrak{sp}(2)} \cong \bbZ_2$.
Therefore, ${\bf c}_1$ itself projects onto an order 2 element in $\Lambda_\text{cw}^\fkg / \Lambda_\text{cr}^\fkg = Z(SU(2) \times SU(2) \times SU(4) \times SU(4) \times Sp(2)) = \bbZ_2 \times \bbZ_2 \times \bbZ_4 \times \bbZ_4 \times \bbZ_2$.
Moreover, this analysis shows that this element must be
\begin{align}
    z({\bf c}_1) = (0,1,2,0,1) \in \bbZ_2 \times \bbZ_2 \times \bbZ_4 \times \bbZ_4 \times \bbZ_2 \, .
\end{align}
An analogous argument shows that ${\bf c}_2$ also projects onto an order 2 element in $Z(\widetilde{G})$, given by
\begin{align}
    z({\bf c}_2) = (1,0,0,2,1) \in \bbZ_2 \times \bbZ_2 \times \bbZ_4 \times \bbZ_4 \times \bbZ_2 \, .
\end{align}
So the global structure of the non-Abelian gauge group $G$ is:
\begin{equation}
    G = \frac{SU(2)^2 \times SU(4)^2 \times Sp(2)}{\bbZ^{(1)}_2 \times \bbZ^{(2)}_2} \, ,
\end{equation}
where the embedding of each $\bbZ_2^{(i)}$ into $Z(\widetilde{G})$ is given by $z({\bf c}_i)$.
Once more, notice the importance of the dual momentum lattice in determining the global gauge group.
Neither ${\bf c}_1$ nor ${\bf c}_2$ are elements of $\Lambda_M$, since $l_2 = 1 \notin 2\bbZ$, so just inspecting points in $\Lambda_M$ would not have yielded this result.
However, ${\bf c}_1 + {\bf c}_2 \in \Lambda_M$, from which one might be tempted to deduce that $\pi_1(G) = \bbZ_2$, which is the diagonal $\bbZ_2 \subset \bbZ^{(1)}_2 \times \bbZ^{(2)}_2$.
Note that this $\bbZ_2$ embeds trivially into $Z(Sp(2))$.

We can explicitly verify, from the generators $z({\bf c}_i)$, that the ${\cal Z} =\bbZ^{(1)}_2 \times \bbZ^{(2)}_2$ 1-form symmetry is free of the anomaly \cite{Cvetic:2020kuw}.
Indeed, we will prove momentarily that this is guaranteed for the non-Abelian gauge group topology of any 8d CHL vacua.

From the lattice embedding, we can also determine the gauge group structure involving the $U(1)$s.
Two generators of $\Lambda_M^*$ that are not contained in $\Lambda_M^* \cap E$ are
\begin{align}
    \begin{split}
    {\bf c}_3 &= (0, 0, 0, 0; \tfrac{1}{2}, -\tfrac{1}{2}, -\tfrac{1}{2}, -\tfrac{1}{2}, \tfrac{1}{2}, -\tfrac{1}{2}, -\tfrac{1}{2}, -\tfrac{1}{2})= \tfrac{1}{4}\boldsymbol\xi_1 + \overline{\bf w}_2 + \overline{\bf w}_3 + \overline{\bf w}_{7} \, , \\
    {\bf c}_4 &= (1, 2, -1, -1; 0, 0, -1, 0, 1, 1, 1, -2)= \tfrac{1}{4}\boldsymbol\xi_2 + \overline{\bf w}_1 + \overline{\bf w}_4 + \overline{\bf w}_8 \, .
    \end{split}
\end{align}
In $\Lambda_\text{cc}' = \pi_E (\Lambda_M^*) \subset \Lambda_\text{cw}^\fkg$, we then have $\pi_E ({\bf c}_3) = \overline{\bf w}_2 + \overline{\bf w}_3 + \overline{\bf w}_{7}$ and $\pi_E({\bf c}_4) = \overline{\bf w}_1 + \overline{\bf w}_4 + \overline{\bf w}_8$, whose equivalence class in $Z(\widetilde{G}) = \bbZ_2 \times \bbZ_2 \times \bbZ_4 \times \bbZ_4 \times \bbZ_2$ are
\begin{equation}
    z({\bf c}_3) = (0, 1, 1, 2, 0)  \, ,\quad z({\bf c}_4) = (1, 0, 2, 1, 0) \, ,
\end{equation}
which each generate a $\bbZ_4$ subgroup.
The first $\bbZ_4$, generated by ${\bf c}_3$, is a subgroup of the $U(1)$ generated by $\boldsymbol\xi_1$, whereas the second $\bbZ_4$ generated by ${\bf c}_4$ is in the $U(1)$ associated with $\boldsymbol\xi_2$.
So, in summary, the global form of the full gauge group is
\begin{equation}
    \frac{[(SU(2)^2 \times SU(4)^2 \times Sp(2))/(\bbZ_2 \times \bbZ_2)] \times U(1)^2}{\bbZ_4 \times \bbZ_4} \, .
\end{equation}

\subsection{Absence of 1-Form Anomalies}
\label{subsec:1-form_anomalies}

A non-trivial global structure $G = \widetilde{G} / {\cal Z}$ for the non-Abelian gauge group can be interpreted as having gauged the subgroup ${\cal Z}$ of the $Z(\widetilde{G})$ 1-form center symmetry of the simply-connected group $\widetilde{G}$ \cite{Gaiotto:2014kfa}.
In supergravity theories of dimension five or higher, such a gauging may be obstructed due to a mixed anomaly involving the large gauge transformations of the tensor field in the supergravity multiplet \cite{Apruzzi:2020zot,Cvetic:2020kuw,BenettiGenolini:2020doj}.
For 8d ${\cal N}=1$ theories, this obstruction can be quantified as follows \cite{Cvetic:2020kuw}.
Let $\widetilde{G} = \prod_i \widetilde{G}_i$, where $\widetilde{G}_i$ are simple factors, with $Z(\widetilde{G}_i) \cong \bbZ_{n_i}$, or $Z(\widetilde{G}_i) \cong \bbZ_2 \times \bbZ_2$ for $\widetilde{G}_i = Spin(4N_i)$.
Then a generator $z$ of ${\cal Z} \subset \prod_i Z(\widetilde{G}_i)$ is specified by a tuple $(k_i)$, where $k_i \mod n_i \in \bbZ_{n_i}$.\footnote{For $\widetilde{G}_i = Spin(4N_i)$ with $Z(\widetilde{G}_i) \cong \bbZ_2 \times \bbZ_2$, we would have two integers $k_i^{(1)}$ and $k_i^{(2)}$ modulo 2 specifying the embedding of $z$ into $Z(\widetilde{G}_i)$.}
The absence of the anomaly that would obstruct the gauging of ${\cal Z}$ requires that for any generator $z \simeq (k_i)$, we have
\begin{align}\label{eq:1-form_anomaly_condition_general}
    \sum_i m_i \, \alpha_{\widetilde{G}_i} \, k_i^2 = 0 \mod \bbZ \, ,
\end{align}
where $m_i$ is the Kac-Moody level of the worldsheet current algebra realization of $\widetilde{G}_i$.
The non-triviality of this condition is due to the fractional numbers $\alpha_{\widetilde{G}_i}$; for $\widetilde{G}$ with non-trivial $Z(\widetilde{G})$ that can appear in 8d supergravity, these are \cite{Cordova:2019uob}:\footnote{For $\widetilde{G} = Spin(4N)$ with $Z(\widetilde{G}) = \bbZ_2 \times \bbZ_2$, there are two inequivalent anomaly coefficients, $(\alpha^{(1)}, \alpha^{(2)})$.
The first is the same for both generators $(1,0)$ and $(0,1)$ of each $\bbZ_2$ factor; the second coefficient is associated with the generator $(1,1)$ of the diagonal $\bbZ_2$ subgroup.
In this identification, the (co-)spinor representation is charged under $(1,0)$ and $(0,1)$, respectively; hence, both are charged under $(1,1)$.
The vector representation is charged under both $(1,0)$ and $(0,1)$, but invariant under $(1,1)$.
\label{footnote:Spin_1}}
\begin{align}\label{eq:list_alpha}
    \renewcommand{\arraystretch}{1.2}
    \begin{array}{c||c|c|c|c|c|c}
        \widetilde{G} & SU(N) & Spin(4N+2) & Spin(4N) & E_6 & E_7 & Sp(N)\\ \hline 
        \alpha_{\widetilde{G}} & \frac{N-1}{2N} & \frac{2N+1}{8} & \left(\frac{N}{4}, \frac12\right) & \frac{2}{3} & \frac34 & \frac{N}{4}
    \end{array}
\end{align}
In the following, we show that for any non-Abelian gauge group $G = \widetilde{G}/{\cal Z}$ realized via lattice embeddings into the Narain or the Mikhailov lattice, as described above, \eqref{eq:1-form_anomaly_condition_general} is satisfied.

To do so, we first recall that any generator $z \simeq (k_1, k_2, ...) \in {\cal Z} = \pi_1(G)$ may be represented by a cocharacter vector ${\bf c} = \imath(\overline{\bf v}_c) \in \Lambda_\text{cc}^G = \Lambda_S^* \cap \imath(E)$.
If $\widetilde{G} = \prod_i \widetilde{G}_i$ with simple factors $\widetilde{G}_i$, then $E = \oplus_i E_i$, where $E_i = \Lambda^{\fkg_i}_\text{r} \otimes \mathbb{R}$, is an orthogonal decomposition of $E$.
So 
\begin{align}\label{eq:splitting_of_cochar_into_coweights}
\begin{split}
    \overline{\bf v}_c = \sum_i \overline{\bf v}^{(i)}_c \, , \quad & \text{with } \, \overline{\bf v}_c^{(i)} \in \Lambda_\text{cw}^{\fkg_i} \, \text{ representing } \, k_i \in \frac{\Lambda_\text{cw}^{\fkg_i}}{\Lambda_\text{cr}^\fkg} = Z(\widetilde{G}_i) \, , \\
    & \text{and } \, (\overline{\bf v}_c^{(i)}, \overline{\bf v}_c^{(j)}) = 0 \, \text{ for } \, i \neq j \, .
\end{split}
\end{align}

The key feature to prove \eqref{eq:1-form_anomaly_condition_general} is that
\begin{align}
\begin{split}
    \langle {\bf c}, {\bf c} \rangle = (\overline{\bf v}_c, \overline{\bf v}_c) = \sum_i (\overline{\bf v}_c^{(i)}, \overline{\bf v}_c^{(i)}) \in \begin{cases}
        2\bbZ & \text{for} \quad {\bf c} \in \Lambda_N^* = \Lambda_N \, , \\
        \bbZ & \text{for} \quad {\bf c} \in \Lambda_M^* \, .
    \end{cases}
\end{split}
\end{align}

Then, to prove \eqref{eq:1-form_anomaly_condition_general} for heterotic vacua (i.e., ${\bf c} \in \Lambda_N$), which only allows ADE-type groups $\widetilde{G}_i$ with $m_i = 1$, we need to show that, for any $\overline{\bf v}_c^{(i)} \in \Lambda_\text{cw}^{\fkg_i} \subset E_i$ which represents $k_i \in Z(\widetilde{G}_i)$, its length square satisfies $(\overline{\bf v}_c^{(i)}, \overline{\bf v}_c^{(i)}) = 2 \alpha_{\widetilde{G}_i} k_i^2 \mod \bbZ$.
For CHL vacua with ${\bf c} \in \Lambda_M^*$, we need $(\overline{\bf v}_c^{(i)}, \overline{\bf v}_c^{(i)}) = 2 \alpha_{\widetilde{G}_i} k_i^2 \mod \bbZ$ for ADE-type $\widetilde{G}_i$ at level 2, and $(\overline{\bf v}_c^{(i)}, \overline{\bf v}_c^{(i)}) = \alpha_{\widetilde{G}_i} k_i^2 \mod \bbZ$ for $\widetilde{G}_i = Sp(N_i)$ at level 1.

For ADE-groups, this simplifies due to $\Lambda_\text{r}^\fkg = \Lambda_\text{cr}^\fkg$ being an even self-dual lattice.
In this case, $Z(\widetilde{G}) = \Lambda_\text{cw}^\fkg / \Lambda_\text{cr}^\fkg = (\Lambda_\text{r}^\fkg)^* / \Lambda_\text{cr}^\fkg = (\Lambda_\text{r}^\fkg)^* / \Lambda_\text{r}^\fkg$ is the so-called discriminant group of $\Lambda_\text{r}^\fkg$ (see \cite{Nikulin_1980} for more details).
Via the pairing on $\Lambda_\text{r}^\fkg \otimes \mathbb{R}$, one can use
\begin{align}
\begin{split}
    & \tfrac12 (\overline{\bf w} + \boldsymbol{\alpha}, \overline{\bf w} + \boldsymbol{\alpha}) = \tfrac12 (\overline{\bf w}, \overline{\bf w}) + (\overline{\bf w}, \boldsymbol{\alpha}) + \tfrac12 (\boldsymbol{\alpha},\boldsymbol{\alpha}) = \tfrac12 (\overline{\bf w}, \overline{\bf w})  \mod \bbZ \, , \\
    & \text{for} \quad \boldsymbol{\alpha} \in \Lambda_\text{r}^\fkg \quad \text{and} \quad \overline{\bf w} \in \Lambda_\text{cw}^\fkg = (\Lambda_\text{r}^\fkg)^* \, ,
\end{split}
\end{align}
to define a quadratic form $q: Z(\widetilde{G}) \rightarrow \mathbb{Q}/\bbZ$, which is a quadratic refinement of the so-called discriminant pairing on $Z(\widetilde{G})$.
Then, if the vector $\overline{\bf v}_c^{(i)} \in \Lambda_\text{cw}^{\fkg_i}$ projects onto $k_i \in (\Lambda_\text{r}^{\fkg_i})^* / \Lambda_\text{r}^{\fkg_i} = Z(\widetilde{G}_i)$, we evidently have $(\overline{\bf v}_c^{(i)}, \overline{\bf v}_c^{(i)}) = 2 q(k_i) \mod \bbZ$.
The upshot of this detour is that the discriminant form of ADE root lattices and its quadratic refinements are well-known (see, e.g., \cite{shimada_k3}), and given by
\begin{align}
    \begin{aligned}
        \mathfrak{su}(N) & : \quad Z(\widetilde{G}) = \bbZ_N\, , \quad && q(k) = k^2 \cdot \tfrac{N-1}{2N} = k^2\, \alpha_{SU(N)}  \, , \\
        \mathfrak{so}(4N+2) & : \quad Z(\widetilde{G}) = \bbZ_4 \, , \quad && q(k) = k^2 \cdot \tfrac{2N+1}{8} = k^2 \, \alpha_{Spin(4N+2)} \, , \\
        \mathfrak{so}(4N) & : \quad Z(\widetilde{G}) = \bbZ_2 \times \bbZ_2 \, , \quad && q( k_1, k_2) = \tfrac{N}{4} (k_1^2 + k_2^2) + \tfrac{N-1}{2} k_1 k_2 \, , \\ 
        \mathfrak{e}_6 &: \quad Z(\widetilde{G}) = \bbZ_3 , \quad && q(k) = k^2 \cdot \tfrac23 = k^2 \, \alpha_{E_6}  \, , \\
        \mathfrak{e}_7 &: \quad Z(\widetilde{G}) = \bbZ_2 , \quad && q(a,b) = k^2 \cdot \tfrac34 = k^2 \, \alpha_{E_7} \, .
    \end{aligned}
\end{align}
Hence, for all simple ADE-type $\widetilde{G}_i$, the quadratic form gives $(\overline{\bf v}_c, \overline{\bf v}_c) = 2q(k_i) = 2k_i^2 \alpha_{\widetilde{G}_i}$, as required to show \eqref{eq:1-form_anomaly_condition_general} for both heterotic and CHL vacua.\footnote{For $\mathfrak{so}(4N)$, the generators $\vec{k} = (1,0), (0,1) \in \bbZ_2 \times \bbZ_2$ both satisfy $q(\vec{k}) = \tfrac{N}{4} = \alpha_{Spin(4N)}^{(1)} \mod \bbZ$, and $\vec{k}=(1,1)$ satisfies $q(k) = N - \tfrac12 = \tfrac12 \mod \bbZ = \alpha^{(2)}_{Spin(4N)}$.
This agrees with the mixed 1-form anomalies with the individual $\bbZ_2$ subgroups, see footnote \ref{footnote:Spin_1}.
}

For $\widetilde{G}_i = Sp(N_i)$, $Z(\widetilde{G}_i) = \bbZ_2$ is no longer the discriminant group of the root lattice, since $(\Lambda_\text{r}^\mathfrak{sp})^* \neq \Lambda_\text{cr}^\mathfrak{sp}$.
So we need to find an explicit representation of $k=1 \in \bbZ_2 = \Lambda_\text{cw}^{\mathfrak{sp}} / \Lambda_\text{cr}^{\mathfrak{sp}}$ in terms of a coweight $\overline{\bf v}$, and compute its length squared.
One way to represent the simple roots of $Sp(N)$ inside $E \cong \mathbb{R}^{N}$, with standard basis $\{\boldsymbol{e}_m\}$, is $\boldsymbol\mu_m = \boldsymbol{e}_m - \boldsymbol{e}_{m+1}$ for $m<N$, and $\boldsymbol\mu_{N} = 2 \boldsymbol{e}_{N}$ (see, e.g., \cite{Hall2015}).
Then, a basis $\overline{\bf w}_l$ for $\Lambda_\text{cw}^{\mathfrak{sp}(N)} = (\Lambda_\text{r}^{\mathfrak{sp}(N)})^*$, which is the dual basis of $\{\boldsymbol{\mu}_m\}$, i.e., $(\overline{\bf w}_l, \boldsymbol{\mu}_m) = \delta_{lm}$, is given by
\begin{align}
    \overline{\bf w}_l = \sum_m (M^{-1})_{l m} \, \boldsymbol{\mu}_m \quad \text{with} \quad M_{l m} = (\boldsymbol{\mu}_l, \boldsymbol{\mu}_m) = \begin{pmatrix}
        2 & -1 & 0 & \ldots & 0 \\
        -1 & 2 & -1 & \ddots & 0 \\
         & \ddots & \ddots & \ddots \\
        0 & \ldots & -1 & 2 & -2 \\
        0 & \ldots & 0 & -2 & 4
    \end{pmatrix}.
\end{align}
The inverse is
\begin{align}
\begin{split}
    (M^{-1})_{k m} & = \min(k, m) \, , \quad k,m < N \, , \\
    (M^{-1})_{Nm} & = (M^{-1})_{mN} = \tfrac{m}{2} \, , \quad m < N \, , \quad (M^{-1})_{NN} = \tfrac{N}{4} \, .
\end{split}
\end{align}
Now, because the coroot lattice $\Lambda_\text{cr}^{\mathfrak{sp}(N)}$ is spanned by $\boldsymbol{\mu}^\vee_m = \boldsymbol{\mu}_m$ for $m<N$, and $\boldsymbol{\mu}^\vee_N = \tfrac12 \boldsymbol{\mu}_N = \boldsymbol{e}_N$, the coweight basis vectors $\overline{\bf w}_k = \sum_m (M^{-1})_{km} \boldsymbol{\mu}_m$ with $k<N$ are actually integer vectors in $\Lambda_\text{cr}^{\mathfrak{sp}(N)}$, and hence represent $0 \in Z(Sp(N)) = \Lambda_\text{cw}^{\mathfrak{sp}(N)} / \Lambda_\text{cr}^{\mathfrak{sp}(N)}$.
The non-trivial element $1 \in \bbZ_2 \cong Z(Sp(N))$ must therefore be the equivalence class of $\overline{\bf w}_N = \sum_{m=1}^{N-1} \tfrac{m}{2} \boldsymbol{\mu}^\vee_m + \tfrac{N}{2} \boldsymbol{\mu}^\vee_N$.
Then, one can explicitly compute that
\begin{align}
\begin{split}
    (\overline{\bf w}_N, \overline{\bf w}_N) = \tfrac{N^2}{2}-\tfrac{N}{4} - 2N+2 &= \tfrac{N^2}{2} - \tfrac{N}{4} - \underbrace{\tfrac{N(N-1)}{2}}_{\in \bbZ \, \, \forall N} \! \! \mod \bbZ  \\
    &= \tfrac{N}{4} \! \! \mod \bbZ  \\
    &= \alpha_{Sp(N)} \mod \bbZ\, ,
\end{split}
\end{align}
which indeed is the form needed to prove \eqref{eq:1-form_anomaly_condition_general} for CHL vacua with $\mathfrak{sp}$ gauge factors.

\section{CHL Gauge Groups from Heterotic Models}
\label{sec:3}

In this section, we show how we can recover the data $({\cal Z}, {\cal Z}')$ about the gauge group topology (cf.~\eqref{eq:global_gauge_group_general}) of any 8d CHL vacua from the corresponding data of an 8d heterotic configuration.
Physically, this is based on the duality between CHL vacua and heterotic compactifications ``without vector structure'' \cite{Witten:1997bs}, or, equivalently F-theory with O7$_+$-planes encoded in ``frozen'' singularities \cite{Tachikawa:2015wka,Bhardwaj:2018jgp}.
In either of these duality frames, an 8d CHL vacuum with non-Abelian gauge algebra $\fkg = \mathfrak{sp}(n) \oplus \fkh$, with $\fkh$ of ADE-type, arises from a heterotic or F-theory model with gauge algebra $\fkg_\text{het} = \mathfrak{so}(16+2n) \oplus \fkh$ (see also \cite{Hamada:2021bbz}).
Indeed, our CHL example in Section \ref{subsec:CHL_example} with $\fkg = \mathfrak{sp}(2) \oplus \mathfrak{su}(4)^2 \oplus \mathfrak{su}(2)^2$ can be obtained from the heterotic example with $\fkg_\text{het} = \mathfrak{so}(20) \oplus \mathfrak{su}(4)^2 \oplus \mathfrak{su}(2)^2$, whose global structure we compute in Appendix \ref{app:heterotic_example}.
A direct comparison shows that both examples have the same data, $({\cal Z}, {\cal Z}') = ({\cal Z}_\text{het}, {\cal Z}'_\text{het})$, which specifies the global structure of the gauge group.
Though, in general, these two pairs need not be identical, the identification of the CHL gauge group data $({\cal Z}, {\cal Z}')$ is straightforward to obtain, given the corresponding information about the heterotic/F-theory model.
The information about the latter can be extracted from various sources, e.g., from K3-data \cite{shimada_k3} specifying the F-theory setting, or the lattice embeddings of heterotic models \cite{Font:2020rsk}.
An alternative way is to use string junctions \cite{Guralnik:2001jh}, which will be explored in full detail in an upcoming work \cite{junctions_tba}.

To describe the procedure, let us assume that we have the explicit embedding of the subgroup $({\cal Z}_\text{het}, {\cal Z}_\text{het}')$ into $Z(\widetilde{G}_\text{het}) = Z(Spin(16+2n)) \times Z(\widetilde{H})$, where $\widetilde{G}_\text{het}$ and $\widetilde{H}$ are the simply-connected groups with algebra $\fkg_\text{het}$ and $\fkh$, respectively.
Then, any generator $z = (z_\mathfrak{sp} , z_\fkh) \in Z(Sp(n)) \times Z(\widetilde{H}) = \bbZ_2 \times Z(\widetilde{H})$ of the group ${\cal Z}$ (or ${\cal Z}'$, respectively) specifying the CHL gauge topology arises from a generator $\hat{z} = (\hat{z}_\mathfrak{so} , \hat{z}_\mathfrak{h})  \in Z(Spin(16+2n))\times Z(\widetilde{H})$ of ${\cal Z}_\text{het}$ (or ${\cal Z}'_\text{het}$, respectively), via the map
\begin{align}\label{eq:map_CHL_center_from_het_center}
\begin{split}
    z_\fkh & = \hat{z}_\fkh \in Z(\widetilde{H}) \, , \\
    z_\mathfrak{sp} & = \begin{cases}
        \hat{z}_\mathfrak{so} \mod 2 \, , & \hat{z}_\mathfrak{so} \in \bbZ_4 = Z(Spin(16+2n)) \quad (n \, \text{ odd}) \, , \\
        \hat{z}_\mathfrak{so}^{(1)} + \hat{z}_\mathfrak{so}^{(2)} \mod 2 \, , & \hat{z}_\mathfrak{so} = (\hat{z}_\mathfrak{so}^{(1)}, \hat{z}_\mathfrak{so}^{(2)}) \in \bbZ_2 \times \bbZ_2 = Z(Spin(16+2n)) \quad (n \, \text{ even}) \, .
    \end{cases}
\end{split}
\end{align}

While we will provide a proof of the validity of this map momentarily, it allows us to readily determine the gauge groups of all CHL vacua, given their heterotic ``parent''.
We illustrate this for all maximally enhanced cases (i.e., where the non-Abelian algebra $\fkg$ has the maximally allowed rank of 10) in Table \ref{tab:big_table} in Appendix \ref{app:big_table}.
We find not only instances with ${\cal Z} = \bbZ_2$, but also examples with ${\cal Z} = \bbZ_2 \times \bbZ_2$.
Moreover, as an explicit check of the claims of Section \ref{subsec:1-form_anomalies}, all these center embeddings correspond to anomaly-free 1-form center symmetries.

\subsection*{Proving the Validity of the Map}

The proof of the validity of \eqref{eq:map_CHL_center_from_het_center} proceeds in three steps.
First, we review the embedding of the Mikhailov lattice and its dual into the Narain lattice \cite{Mikhailov:1998si}, and highlight the role of the $\mathfrak{so}(16) \subset \mathfrak{so}(16+2n)$ subalgebra.
Second, we construct the roots and coroots of the CHL gauge algebra $\fkg$ from those of the parent heterotic algebra $\fkg_\text{het}$.
Because $\fkg \supset \mathfrak{sp}(n)$, the coroot lattice of $\fkg$ will no longer be a sublattice of the heterotic configuration.
In the third step, we show that the cocharacter lattice of the CHL configuration is obtained from a suitable projection of the cocharacter lattice of the heterotic model.
Analogously to Section \ref{subsec:1-form_anomalies}, each cocharacter $\hat{\bf c}$ projects onto coweights of each gauge factor, thereby specifying a generator of the fundamental group as embedded into the center of the simply-connected cover.
Then, by identifying the generators of $\Lambda_\text{cw}^\fkg / \Lambda_\text{cr}^\fkg$ in terms of those of $\Lambda_\text{cw}^{\fkg_\text{het}} / \Lambda_\text{cr}^{\fkg_\text{het}}$, we will establish the map \eqref{eq:map_CHL_center_from_het_center}.

\subsubsection*{Finding Mikhailov inside Narain}

As argued in \cite{Mikhailov:1998si}, there is, up to isomorphisms, a unique embedding
\begin{align}\label{eq:unique_D8_embedding}
    \Lambda_\text{r}^{\mathfrak{so}(16)} \equiv \text{D}_8 \hookrightarrow \Gamma_{16} \subset \Gamma_{16} \oplus U \oplus U = \Lambda_N
\end{align}
of the root lattice of $\mathfrak{so}(16)$ into the Narain lattice.
Denoting by $V_M$ the subspace of $V_N := \Lambda_N \otimes \mathbb{R}$ that is orthogonal to $\text{D}_8$, with projection $P_M: V_N \rightarrow V_M$, the Mikhailov lattice $\Lambda_M$ and its dual $\Lambda_M^*$ are found as
\begin{align}\label{eq:mikhailov_inside_narain}
\begin{split}
    P_M(\Lambda_N) & \cong \text{D}_8^* \oplus U \oplus U \cong \Lambda_M^* \, , \\
    \Lambda_N \cap V_M & \cong \text{D}_8 \oplus U \oplus U \cong \Lambda_M \, .
\end{split}
\end{align}

To give an ``intuitive'' argument for this, note that the lattice $\Gamma_{16}$ can be identified with the character lattice of $Spin(32)/\bbZ_2$, i.e., it is generated by the root lattice of $\mathfrak{so}(32)$ together with the weights of the spinor representation ${\bf S}_{\mathfrak{so}(32)}$.
The embedding $\text{D}_8 \hookrightarrow \Gamma_{16}$ then corresponds to the embedding of the roots of an $\mathfrak{so}(16) \subset \mathfrak{so}(32)$ subalgebra.
Conversely, the branching $\mathfrak{so}(32) \supset \mathfrak{so}(16) \oplus \mathfrak{so}(16)$ corresponds to an orthogonal decomposition $\Gamma_{16} \otimes \mathbb{R} = (\text{D}_8 \otimes \mathbb{R}) \oplus (\text{D}_8 \otimes \mathbb{R}) \equiv V_{D_8} \oplus V_{D_8}'$.
We can extend this decomposition to
\begin{align}\label{eq:unique_D8_embedding_vector_space}
\begin{split}
    & V_N = \Lambda_N \otimes \mathbb{R} = V_{D_8} \oplus \underbrace{V_{D_8}' \oplus (U \otimes \mathbb{R}) \oplus (U \otimes \mathbb{R})}_{=:V_M} \, , \\
    & P_M: V_N \rightarrow V_M \, , \quad {\bf s} = {\bf s}_D + {\bf s}_M \mapsto {\bf s}_M \, .
\end{split}
\end{align}
From this, the nature of the two lattices $P_M(V_N)$ and $\Lambda_N \cap V_M$ can be inferred from the group-theoretic decomposition
\begin{align}
\begin{split}
    \mathfrak{so}(32) & \supset \mathfrak{so}(16) \oplus \mathfrak{so}(16) \, ,\\
    {\bf adj}_{\mathfrak{so}(32)} & \rightarrow ({\bf adj}_{\mathfrak{so}(16)}, {\bf 1}) \oplus ({\bf 1}, {\bf adj}_{\text{so}(16)}) \oplus ({\bf V}_{\mathfrak{so}(16)}, {\bf V}_{\mathfrak{so}(16)}) \, ,\\
    {\bf S}_{\mathfrak{so}(32)} & \rightarrow ({\bf S}_{\mathfrak{so}(16)}, {\bf S}_{\mathfrak{so}(16)}) \oplus ({\bf C}_{\mathfrak{so}(16)} , {\bf C}_{\mathfrak{so}(16)}) \, ,
\end{split}
\end{align}
where ${\bf adj}$, ${\bf V}$ and ${\bf C}$ denote the adjoint, vector, and co-spinor representations, respectively.
At the level of lattices, the lack of any non-adjoint representations that are charged under just one of the $\mathfrak{so}(16)$ factors means that the only lattice points in the hyperplane $V'_{D_8}$ correspond to adjoint weights, i.e., $\Gamma_{16} \cap V'_{D_8} \cong \text{D}_8$.\footnote{The symmetry between the two $\text{D}_8$'s is an isomorphism of $\Lambda_N$. The results below would be the same if we swapped their roles in the subsequent discussion.}
However, since the bi-charged representations project onto the (co-)spinors and vectors of each $\mathfrak{so}(16)$, the projection of $\Gamma_{16}$ onto $V'_{D_8}$ is $\Lambda_\text{w}^{\mathfrak{so}(16)} = \left(\Lambda_\text{cr}^{\mathfrak{so}(16)} \right)^* = \left(\Lambda_\text{r}^{\mathfrak{so}(16)} \right)^* = \text{D}_8^*$.
Since the copies of $U$ lattices in \eqref{eq:unique_D8_embedding} and \eqref{eq:unique_D8_embedding_vector_space} are merely spectators in this argument, we find the Mikhailov lattice $\Lambda_M$ and its dual as given in \eqref{eq:mikhailov_inside_narain}.

\subsubsection*{Constructing the CHL (co-)roots}

Since the heterotic gauge algebras $\fkg_\text{het} = \fkh \oplus \mathfrak{so}(16+2n)$, which are of interest to us, contain an $\mathfrak{so}(16+2n)$ algebra, we can identify an $\mathfrak{so}(16) \subset \mathfrak{so}(16+2n)$ subalgebra, whose root lattice may be identified with $\text{D}_8$ in \eqref{eq:unique_D8_embedding}.
By orthogonality \eqref{eq:unique_D8_embedding_vector_space}, the root lattice $\Lambda_\text{r}^\fkh \subset \Lambda_N$ of the ADE-algebra $\fkh$ must then lie in the plane $V_M$, and hence, by \eqref{eq:mikhailov_inside_narain}, $\Lambda_\text{r}^\fkh = \Lambda_\text{cr}^\fkh  \subset \Lambda_M$.

In order to obtain the $\mathfrak{sp}(n)$, first consider a basis for $\Lambda_\text{r}^{\mathfrak{so}(16+2n)}$ formed by the simple roots $\hat{\boldsymbol{\mu}}_i$, $i=1,...,8+n$, of $\mathfrak{so}(16+2n)$, with $\hat{\boldsymbol{\mu}}_{n+7}$ and $\hat{\boldsymbol{\mu}}_{n+8}$ forming the ``branched nodes'' in the $\mathfrak{so}(16+2n)$ Dynkin diagram:
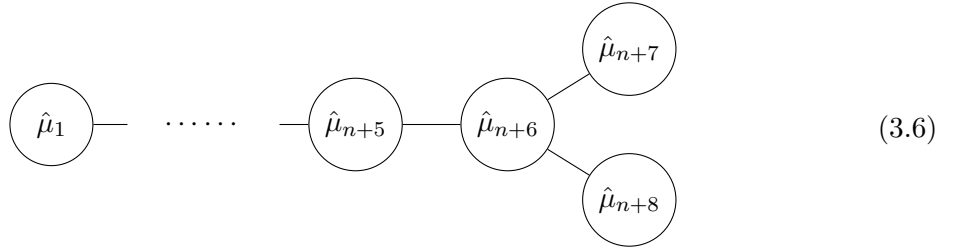
\begin{equation}
\begin{split}
\begin{tikzpicture}
        \node [style=A] (13) at (1, 0) {$\ \hat{\mu}_{1}\ $};
        \node [style=none] (10) at (2, 0) {};
        \node [style=none] (18) at (3, 0) {$\cdots \cdots$};
        \node [style=none] (17) at (4, 0) {};
		\node [style=A] (4) at (5, 0) {$\hat{\mu}_{n + 5}$};
		\node [style=A] (5) at (7, 0) {$\hat{\mu}_{n + 6}$};
		\node [style=A] (6) at (8.6, 1) {$\hat{\mathbf{\mu}}_{n + 7}$};
		\node [style=A] (7) at (8.6, -1) {$\hat{\mu}_{n + 8}$};
		\draw [in=180, out=0] (4) to (5);
		\draw (5) to (6);
		\draw (5) to (7);
		\draw (13) to (10.center);
		\draw (17.center) to (4);
\end{tikzpicture}
\end{split}
\end{equation}
Then, associated with the branching $\mathfrak{so}(16+2n) \supset \mathfrak{so}(16) \oplus \mathfrak{so}(2n)$, the subspace $V_{D_8}$ in \eqref{eq:unique_D8_embedding_vector_space} is spanned by the $\mathfrak{so}(16)$ roots $\{ \hat{\boldsymbol{\mu}}_{n+1}, ..., \hat{\boldsymbol{\mu}}_{n+8}\}$.
Orthogonal to that will be the root lattice $\Lambda_\text{r}^{\mathfrak{so}(2n)}$ of $\mathfrak{so}(2n)$ inside $\text{D}_8 \oplus U \oplus U \cong \Lambda_M$, with simple roots
\begin{align}\label{eq:simple_roots_remainig_so}
    \hat{\boldsymbol{\rho}}_1 = \hat{\boldsymbol{\mu}}_{n-1} \, , \quad \hat{\boldsymbol{\rho}}_2 = \hat{\boldsymbol{\mu}}_{n-2} , \quad ... \quad  , \, \hat{\boldsymbol{\rho}}_{n-1} = \hat{\boldsymbol{\mu}}_{1} \, , \quad \hat{\boldsymbol{\rho}}_n = \hat{\boldsymbol{\mu}}_1 + \sum_{i=2}^{n+6} 2 \hat{\boldsymbol{\mu}}_{i} + \hat{\boldsymbol{\mu}}_{n+7} + \hat{\boldsymbol{\mu}}_{n+8} \, .
\end{align}
At the level of lattices, we have $\Lambda_\text{r}^{\mathfrak{so}(2n)} = \Lambda_\text{r}^{\mathfrak{sp}(n)}$, but the simple roots differ.
In terms of the $\mathfrak{so}(2n)$ roots $\hat{\boldsymbol{\rho}}$, the simple roots $\boldsymbol{\rho}$ of $\mathfrak{sp}(n)$ are \cite{Hall2015}
\begin{align}\label{eq:simple_roots_sp_from_so}
    {\boldsymbol{\rho}}_1 = \hat{\boldsymbol{\rho}}_1 \, , \quad {\boldsymbol{\rho}}_2 = \hat{\boldsymbol{\rho}}_2 \, , \quad ... \quad , \quad {\boldsymbol{\rho}}_{n-1} = \hat{\boldsymbol{\rho}}_{n-1} \, , \quad {\boldsymbol{\rho}}_n = -(\hat{\boldsymbol{\rho}}_{n-1} + \hat{\boldsymbol{\rho}}_{n}) \, ,
\end{align}
where $\boldsymbol{\rho}_n$ is the long root of $\mathfrak{sp}(n)$.
This modifies to coroot lattice $\Lambda_\text{cr}^\mathfrak{so(2n)} \neq \Lambda_\text{cr}^\mathfrak{sp(n)} \subset \Lambda_M^*$, with basis $\boldsymbol{\rho}_i^\vee = \boldsymbol{\rho}_i$ for $i=1,...,n-1$, and $\boldsymbol{\rho}_n^\vee = \tfrac12 \boldsymbol{\rho}_n$.
Under the projection $P_M: V_N \rightarrow V_M$, we have
\begin{align}\label{eq:projection_of_so_roots_as_sp}
\begin{split}
    P_M(\hat{\boldsymbol{\mu}}_i) & = 0 \quad \text{for} \quad i = n+1,...,n+8 \, , \\
    P_M(\hat{\boldsymbol{\mu}}_i) & = \hat{\boldsymbol{\mu}}_i = \boldsymbol{\rho}_{n-i} = \boldsymbol{\rho}_{n-i}^\vee \quad \text{for} \quad i=1,...,n-1 \, , \\
    P_M(\hat{\boldsymbol{\mu}}_n) & = \frac12 P_M \left(\hat{\boldsymbol{\rho}}_n - \hat{\boldsymbol{\mu}}_1 - \sum_{i=2}^{n-1} 2\hat{\boldsymbol{\mu}}_i - \sum_{i=n+1}^{n+6} 2\hat{\boldsymbol{\mu}}_i - \hat{\boldsymbol{\mu}}_{n+7} - \hat{\boldsymbol{\mu}}_{n+8} \right) \\ 
    & = \frac12 \left( \hat{\boldsymbol{\rho}}_n - \hat{\boldsymbol{\rho}}_{n-1} - 2\sum_{i=1}^{n-2} \hat{\boldsymbol{\rho}}_i \right) = - \frac{\boldsymbol{\rho}_n}{2} - \sum_{i=1}^{n-1} \boldsymbol{\rho}_i = - \boldsymbol{\rho}_n^\vee - \sum_{i=1}^{n-1} \boldsymbol{\rho}_i^\vee \, .
\end{split}
\end{align}

Note that, by our working assumption, the $\mathfrak{so}$-roots \eqref{eq:simple_roots_remainig_so} satisfy the masslessness condition for the heterotic string. 
A legitimate question is, then, if the $\mathfrak{sp}$-roots \eqref{eq:simple_roots_sp_from_so} satisfy the analogous conditions of the CHL string, i.e., whether the corresponding CHL vacuum indeed has an $\mathfrak{sp}(n) \oplus \fkh$ gauge symmetry. This is indeed the case, since the embedding above is directly related to the realization of $\mathfrak{sp}$ gauge algebras given in \cite{Mikhailov:1998si}.

\subsubsection*{CHL cocharacters from heterotic cocharacters}

Having identified the (co-)root lattices, we now need to show that every cocharacter of the CHL configuration arises from a cocharacter in the heterotic model.
By defining $\widehat{E} = \Lambda_\text{r}^{\fkg_\text{het}} \otimes \mathbb{R} \subset V_N$ and $E = \Lambda_\text{r}^\fkg \otimes \mathbb{R} \subset V_M$, we first want to show that
\begin{align}\label{eq:projection_commutes_w_cap}
    P_M(\Lambda_N) \cap E = P_M( \Lambda_N \cap \widehat{E})\,.
\end{align}
For this, we use the fact that the branching $\fkg_\text{het} = \mathfrak{so}(16+2n) \oplus \fkh  \supset \mathfrak{so}(16)  \oplus \mathfrak{so}(2n) \oplus \fkh $ induces the orthogonal decomposition
\begin{align}
    \widehat{E} = V_{D_8} \oplus ( \underbrace{\Lambda_\text{r}^{\mathfrak{so}(2n)} \oplus \Lambda_\text{r}^{\fkh}}_{= \Lambda_\text{r}^\fkg}) \otimes \mathbb{R} = V_{D_8} \oplus E \subset V_{D_8} \oplus V_M = V_N\, .
\end{align}
Combining this with the general decomposition \eqref{eq:decomp_V_into_E+F} of $V_N = \widehat{E} \oplus F$, where $F$ is the hyperplane containing the $U(1)$s, \eqref{eq:unique_D8_embedding_vector_space} implies that
\begin{align}\label{eq:V_N_complete_decomp}
    V_M = E \oplus F \subset \underbrace{V_{D_8} \oplus E}_{\widehat{E}} \oplus F = V_N \, .
\end{align}
Notice that, in particular, the number of independent $U(1)$ gauge factors, $r_F = \dim F$, is the same for the CHL and the heterotic vacuum.
So, any ${\bf s} \in V_N$ can be written as ${\bf s} = {\bf s}_D + {\bf s}_E + {\bf s}_F$ with ${\bf s}_D \in V_{D_8}$, ${\bf s}_E \in E$, and ${\bf s}_F \in F$.
Then,
\begin{align}
\begin{split}
    {\bf s} \in P_M(\Lambda_N) \cap E & \quad \Leftrightarrow \quad {\bf s} = {\bf s}_E \in E \ \ \text{and} \ \ \exists \ {\bf s}_D \in V_{D_8} : {\bf s}_E + {\bf s}_D \in \Lambda_N \\
    & \quad \Leftrightarrow \quad \exists \ {\bf s}' = {\bf s}_E + {\bf s}_D \in \Lambda_N \cap \widehat{E} : {\bf s} := {\bf s}_E = P_M({\bf s}') \\
    & \quad \Leftrightarrow \quad {\bf s} \in P_M(\Lambda_N \cap \widehat{E}) \, .
\end{split}
\end{align}
The significance of \eqref{eq:projection_commutes_w_cap} is that we can identify the cocharacter lattice $\Lambda^G_\text{cc}$ of the CHL vacuum as the projection of the heterotic cocharacter lattice $\Lambda^{G_\text{het}}_\text{cc}$ under $P_M$.\footnote{We can also determine the CHL group structure including the $U(1)$s, following \eqref{eq:ab_group_from_string_lat}, from the parent heterotic theory. We will focus on the non-Abelian part, because the relevant data for its group topology are encoded in known K3-data \cite{shimada_k3}.
}
Namely, from the general prescription \eqref{eq:non_ab_group_from_string_lat}, we have
\begin{align}
    \Lambda^G_\text{cc} = \Lambda_M^* \cap E \stackrel{\eqref{eq:mikhailov_inside_narain}}{=} P_M(\Lambda_N) \cap E \stackrel{\eqref{eq:projection_commutes_w_cap}}{=} P_M(\Lambda_N \cap \widehat{E}) = P_M (\Lambda^{G_\text{het}}_\text{cc}) \, .
\end{align}

Analogously, we can infer the CHL cocharacters $\Lambda_\text{cc}'$, that encode the constraints involving the $U(1)$ charges, from the corresponding ones of the heterotic model, $\widehat{\Lambda}_\text{cc}^{'}$.
Namely, at the level of vector spaces, \eqref{eq:V_N_complete_decomp} implies that the projections $\pi_{\widehat{E}}: V_N \rightarrow \widehat{E}$ and $P_M: V_N \rightarrow V_M$ commute, and in fact compose to the projection $\pi_E: V_N \rightarrow E$.
Then, from \eqref{eq:ab_group_from_string_lat}, we have 
\begin{align}
    \Lambda'_{\text{cc}} = \pi_E (\Lambda_M^*) = \pi_E (P_M(\Lambda_N)) = P_M(\pi_E(\Lambda_N)) = P_M (\widehat{\Lambda}'_{\text{cc}}) \, .
\end{align}

In summary, we see that any cocharacter ${\bf c}$ of the CHL vacuum arises as the projection of a heterotic cocharacter $\hat{\bf c}$ under $P_M$.
Any such cocharacter $\hat{\bf c} \in \Lambda_N$ can be written as $\hat{\bf c} = \hat{\bf c}_{\mathfrak{so}(16+2n)} + \hat{\bf c}_{\fkh} + \hat{\bf c}_F$.
If $\hat{\bf c}_F \in F$ is 0, then $\hat{\bf c} \in \Lambda_N \cap \widehat{E} = \Lambda_\text{cc}^{G_\text{het}}$ specifies an element $\hat{z}$ of $\pi_1(G_\text{het}) = {\cal Z}_\text{het} = \Lambda_\text{cc}^{G_\text{het}} / \Lambda_\text{cr}^{\fkg_\text{het}} \subset \Lambda_\text{cw}^{\fkg_\text{het}} / \Lambda_\text{cr}^{\fkg_\text{het}} = Z(\widetilde{G}_\text{het})$.
If $\hat{\bf c}_F \neq 0$, then $\hat{\bf c}$ defines an element $\hat{z}$ of ${\cal Z}'_\text{het} \subset Z(\widetilde{G}_\text{het}) \times U(1)^{r_F}$.
In particular, each component $\hat{\bf c}_\fkh \in \Lambda_\text{cw}^\fkh$ and $\hat{\bf c}_{\mathfrak{so}(16+2n)} \in \Lambda_\text{cw}^{\mathfrak{so}(16+2n)}$ specifies a center element $\hat{z}_{\fkh} \in Z(\widetilde{H})$ and $\hat{z}_\mathfrak{so} \in Z(Spin(16+2n))$, respectively, which is the restriction of $\hat{z}$ to the corresponding center subgroup.

To establish \eqref{eq:map_CHL_center_from_het_center}, all we need to determine is, given the generators of ${\cal Z}_\text{het}$ and ${\cal Z}'_\text{het}$ in terms of $(\hat{z}_\fkh, \hat{z}_\mathfrak{so}) \in Z(\widetilde{H}) \times Z(Spin(16+2n)) = Z(\widetilde{G}_\text{het})$, what the corresponding center element $z = (z_\fkh, z_{\mathfrak{sp}}) \in Z(\widetilde{H}) \times Z(Sp(n)) = Z(\widetilde{G})$ is.
For $z_\fkh$, this is easy to answer.
Since the coroot lattice $\Lambda_\text{cr}^\fkh$ remains invariant when passing from the heterotic to the CHL model, the component $P_M(\hat{\bf z}_\fkh) = \hat{\bf z}_\fkh$ defines the same element $z_{\fkh} = \hat{z}_\fkh \in Z(\widetilde{H}) = \Lambda_\text{cw}^\fkh / \Lambda_\text{cr}^\fkh$.

However, the same does not hold for $z_\mathfrak{sp}$, since in the CHL vacuum, we have to compare $P_M(\hat{\bf c}_{\mathfrak{so}(16+2n)})$ to the coroots of $\mathfrak{sp}(n)$, rather than those of $\mathfrak{so}(2n) \subset \mathfrak{so}(16+2n)$.
To this end, we first construct the coweights $\widehat{\overline{\bf w}}$ of $\mathfrak{so}(16+2n)$ which represent $Z(Spin(16+2n)) = \Lambda_\text{cw}^{\mathfrak{so}(16+2n)} / \Lambda_\text{cr}^{\mathfrak{so}(16+2n)}$.
As duals of the roots $\hat{\boldsymbol{\mu}}_i$, $i=1,...,8+n$, a basis for these are given by
\begin{align}
    \widehat{\overline{\bf w}}_l = (C^{-1})_{l i} \hat{\boldsymbol{\mu}}_i \quad \text{with} \quad C_{il} = (\hat{\boldsymbol{\mu}}_i, \hat{\boldsymbol{\mu}}_l) = \begin{pmatrix}
        2 & -1 & 0 & \ldots & \ldots & 0 \\
        -1 & 2 & \ddots & \ddots & & \\
        & \ddots & \ddots & \ddots & \\
        0 & \ldots & -1 & 2 & -1 & -1 \\
        0 & \ldots & 0 & -1 & 2 & 0 \\
        0 & \ldots & 0 & -1 & 0 & 2
    \end{pmatrix}
\end{align}
The inverse of the Cartan matrix $C$ of $\mathfrak{so}(16+2n)$ is (see, e.g., \cite{2017arXiv171101294W})
\begin{align}\label{eq:inverse_so_cartan_matrix}
\begin{split}
    (C^{-1})_{ij} & = (C^{-1})_{ji} = \min(i,j) \quad \text{for} \quad i,j < n+6 \, , \\
    (C^{-1})_{n+7,j} & = (C^{-1})_{j,n+7} = (C^{-1})_{n+8,j} = (C^{-1})_{j,n+8} = \tfrac{j}{2} \quad \text{for} \quad j < n+6 \, , \\
    (C^{-1})_{n+7,n+8} & = (C^{-1})_{n+8,n+7} = \tfrac{n+6}{4} \, , \quad (C^{-1})_{n+7,n+7} = (C^{-1})_{n+8,n+8} = \tfrac{n+8}{4} \, .
\end{split}
\end{align}
Since we are ultimately interested in the equivalence classes of $\mathfrak{sp}(n)$-coweights ${\bf c}_{\mathfrak{sp}}$ in $\bbZ_2 \cong \Lambda_\text{cw}^{\mathfrak{sp}(n)} / \Lambda_\text{cr}^{\mathfrak{sp}(n)}$, we use \eqref{eq:simple_roots_sp_from_so} to compute, for later convenience,
\begin{align}\label{eq:projection_so_coweights}
\begin{split}
    P_M(\widehat{\overline{\bf w}}_{n+7}) & = \sum_{j=1}^{n+8} (C^{-1})_{n+7, j} \, P_M(\hat{\boldsymbol{\mu}}_j) \\
    & = \sum_{j=1}^{n-1} (C^{-1})_{n+7,j} \, \boldsymbol{\rho}_{n-j}^\vee - (C^{-1})_{n+7,n} \left( \boldsymbol{\rho}_n^\vee + \sum_{j=1}^{n-1} \boldsymbol{\rho}_j^\vee \right) \\
    & = \sum_{j=1}^{n-1} (C^{-1})_{n+8,j} \, \boldsymbol{\rho}_{n-j}^\vee - (C^{-1})_{n+8,n} \left( \boldsymbol{\rho}_n^\vee + \sum_{j=1}^{n-1} \boldsymbol{\rho}_j^\vee \right) = P_M(\widehat{\overline{\bf w}}_{n+8}) \\
    & = \sum_{j=1}^{n-1} \frac{j-n}{2} \boldsymbol{\rho}_j^\vee - \frac{n}{2} \boldsymbol{\rho}_n^\vee \, .
\end{split}
\end{align}
Clearly, for any $n\geq 1$, at least one of the summands has a fractional coefficient.
And since $2P_M(\widehat{\overline{\bf w}}_{n+7}) = 2P_M(\widehat{\overline{\bf w}}_{n+8}) \in \Lambda_\text{cr}^{\mathfrak{sp}(n)}$, this means that $P_M(\widehat{\overline{\bf w}}_{n+7}) = P_M(\widehat{\overline{\bf w}}_{n+8})$ map to $1 \in \bbZ_2 \cong \Lambda_\text{cw}^{\mathfrak{sp}(n)} / \Lambda_\text{cr}^{\mathfrak{sp}(n)}$.
Moreover, since $\hat{\boldsymbol{\mu}}^\vee_i = \hat{\boldsymbol{\mu}}_i$, we can easily verify that
\begin{align}
\begin{split}\label{eq:relation_coweights_so}
    & \widehat{\overline{\bf w}}_{n+7} + \widehat{\overline{\bf w}}_{n+8} = \sum_{j=1}^{n+8} ((C^{-1})_{n+7,j} + (C^{-1})_{n+8,j}) \hat{\boldsymbol{\mu}}_j \\
    = & \sum_{j=1}^{n+6} j \, \hat{\boldsymbol{\mu}}_j + \frac{2n+14}{4} ( \hat{\boldsymbol{\mu}}_{n+7} + \hat{\boldsymbol{\mu}}_{n+8}) \\
    = & \frac{n+1}{2} ( \hat{\boldsymbol{\mu}}_{n+7}^\vee + \hat{\boldsymbol{\mu}}_{n+8}^\vee) \mod \Lambda_\text{cr}^{\mathfrak{so}(16+2n)} \, .
\end{split}
\end{align}
Now it is instructive to differentiate between even and odd $n$.

For odd $n$, for which we know $\Lambda_\text{cw}^{\mathfrak{so}(16+2n)} / \Lambda_\text{cr}^{\mathfrak{so}(16+2n)} \cong \bbZ_4$, the above equation is $\widehat{\overline{\bf w}}_{n+7} + \widehat{\overline{\bf w}}_{n+8} = 0 \mod \Lambda_\text{cr}^{\mathfrak{so}(16+2n)}$.
At the same time, since $2(n+6)$ and $2(n+8)$ cannot be divisible by 4 with odd $n$, both $\widehat{\overline{\bf w}}_{n+7}$ and $\widehat{\overline{\bf w}}_{n+8}$ are order 4 elements modulo $\Lambda_\text{cr}^{\mathfrak{so}(16+2n)}$.
The order 2 element in $\Lambda_\text{cw}^{\mathfrak{so}(16+2n)} / \Lambda_\text{cr}^{\mathfrak{so}(16+2n)}$ is then represented by $2 \widehat{\overline{\bf w}}_{n+7} = 2 \widehat{\overline{\bf w}}_{n+8} \mod \Lambda_\text{cr}^{\mathfrak{so}(16+2n)} = \widehat{\overline{\bf w}}_{2j-1} \mod \Lambda_\text{cr}^{\mathfrak{so}(16+2n)}$ for $1\leq j \leq n+6$ (as also evident from \eqref{eq:inverse_so_cartan_matrix}).
Then, if $\hat{\bf c}_{\mathfrak{so}(16+2n)}$ projects onto an order 4 element in $\Lambda_\text{cw}^{\mathfrak{so}(16+2n)} / \Lambda_\text{cr}^{\mathfrak{so}(16+2n)} \cong \bbZ_4$, it must be in the same equivalence class as either $\widehat{\overline{\bf w}}_{n+7}$ or $\widehat{\overline{\bf w}}_{n+8}$, which by \eqref{eq:projection_of_so_roots_as_sp} both map onto the order 2 element in $\Lambda_\text{cw}^{\mathfrak{sp}(n)} / \Lambda_\text{cr}^{\mathfrak{sp}(n)} \cong \bbZ_2$, confirming \eqref{eq:map_CHL_center_from_het_center} for odd $n$.

For even $n$, we see from \eqref{eq:inverse_so_cartan_matrix} and \eqref{eq:relation_coweights_so} that, in the quotient $\Lambda_\text{cw}^{\mathfrak{so}(16+2n)} / \Lambda_\text{cr}^{\mathfrak{so}(16+2n)} \cong \bbZ_2 \times \bbZ_2$, the equivalence classes of $\widehat{\overline{\bf w}}_{n+7}$, $\widehat{\overline{\bf w}}_{n+8}$, and $\widehat{\overline{\bf w}}_{n+7} + \widehat{\overline{\bf w}}_{n+8}$ all define order 2 elements.
This means that $\widehat{\overline{\bf w}}_{n+7}$ and $\widehat{\overline{\bf w}}_{n+8}$ represent the generators $(1,0)$ and $(0,1)$, respectively, while $\widehat{\overline{\bf w}}_{n+7} + \widehat{\overline{\bf w}}_{n+8}$ represents $(1,1)$.
Again, since $P_M(\widehat{\overline{\bf w}}_{n+7}) = P_M(\widehat{\overline{\bf w}}_{n+8})$ map to $1 \in \bbZ_2 = Z(Sp(n))$, this confirms \eqref{eq:map_CHL_center_from_het_center} for even $n$.

\section{Summary and Outlook}
\label{sec:conclusions}

In this work, we have presented an explicit identification of the gauge group topology
\begin{align}
    \frac{[\widetilde{G}/{\cal Z}] \times U(1)^{r_F}}{{\cal Z}'}
\end{align}
of 8d ${\cal N}=1$ compactifications of heterotic and CHL string theories, based on the embedding of the root lattice $\Lambda_\text{r}^\fkg$ of the non-Abelian gauge algebra $\fkg$ (with simply-connected cover $\widetilde{G}$) into the momentum lattice $\Lambda_S$ of string states.
For rank 20 theories, this agrees with known results from the heterotic \cite{Font:2020rsk} or the F-theory duality frame \cite{Aspinwall:1998xj,Guralnik:2001jh,shimada_k3,junctions_tba}.
For CHL vacua, we have highlighted the necessity to distinguish between $\Lambda_S$ and its dual, as well as between the root $\Lambda_\text{r}^\fkg$ and coroot lattice $\Lambda_\text{cr}^\fkg$.
If this is taken into account, the resulting non-Abelian gauge group topology $G = \widetilde{G}/{\cal Z}$ is guaranteed to have no anomalies for the corresponding ${\cal Z} \subset Z(\widetilde{G})$ 1-form symmetry \cite{Cvetic:2020kuw}.
This can be verified explicitly for all 61 maximally enhanced CHL vacua, for which we have compiled the non-Abelian gauge group topology ${\cal Z}$ in Appendix \ref{app:big_table}.

We have also demonstrated in an explicit example how to compute the subgroup ${\cal Z}' \subset Z(\widetilde{G})$, which is identified with a subgroup of the Abelian gauge factor $U(1)^{r_F}$.
As we have argued for in Section \ref{sec:3}, the global gauge group structure $({\cal Z}, {\cal Z}')$ of any CHL vacuum can be in principle inferred from the corresponding data $({\cal Z}_\text{het}, {\cal Z}'_\text{het})$ of a parent heterotic model.
While ${\cal Z}_\text{het}$ can be readily obtained from existing data (e.g., from \cite{Font:2020rsk}), a comprehensive list of the part ${\cal Z}_\text{het}'$ involving the $U(1)$s will be presented in an upcoming work \cite{junctions_tba}, from which we can then also classify ${\cal Z}'$ for CHL vacua.
There, we will also extend the analysis to include 8d ${\cal N}=1$ theories with gauge rank 4 \cite{Dabholkar:1996pc,Witten:1997bs,Aharony:2007du}.
Additionally, it should be straightforward to apply the machinery to 7d heterotic compactifications \cite{Fraiman:2021soq}.

Another interesting direction would be to understand the results about the global gauge group structures involving geometrically engineered $\mathfrak{sp}$ gauge symmetries in the language of higher-form symmetries \cite{Gaiotto:2014kfa}.
This would require a refinement of the framework of \cite{Morrison:2020ool,Albertini:2020mdx} to M-theory compactifications with frozen singularities \cite{Witten:1997bs,Tachikawa:2015wka,Bhardwaj:2018jgp}.
Furthermore, it would be interesting to reproduce the $\mathfrak{sp}(n)$-contribution to the mixed 1-form anomalies from a dimensional reduction of the M-theory Chern--Simons term in the presence of boundary fluxes which encode the 1-form symmetry background \cite{Cvetic:2021sxm}.
Lastly, having a complete catalog of gauge group topology including the $U(1)$s could provide a guideline to formulate field-theoretic constraints on allowed topologies ${\cal Z}'$, in similar fashion to \cite{Cvetic:2020kuw,Montero:2020icj}.

\section*{Acknowledgments}

We thank Miguel Montero for valuable discussions.
M.C.~and H.Y.Z.~are supported in part by DOE Award No.~DE-SC013528Y.
M.C.~further acknowledges support by the Simons Foundation Collaboration Grant \#724069 on ``Special Holonomy in Geometry, Analysis and Physics'', the Slovenian Research Agency (ARRS Grant No.~P1-0306), and the Fay R.~and Eugene L.~Langberg Endowed Chair.

\begin{appendix}

\section{A Heterotic Case Study}
\label{app:heterotic_example}

In this appendix, we study the global gauge group structure of a rank 20 heterotic model, with $\fkg = \mathfrak{su}(2)^2 \oplus \mathfrak{su}(4)^2 \oplus \mathfrak{so}(20)$.

We choose a presentation of the Narain lattice $\Lambda_N$ and its vector space $V_N = \Lambda_N \otimes \mathbb{R}$ as
\begin{align}
    V_N \ni {\bf v}^{(\ell)} =  (l^{(\ell)}_1, l^{(\ell)}_2, n^{(\ell)}_1, n^{(\ell)}_2; s^{(\ell)}_1, \dots, s^{(\ell)}_{16} ) \, ,
\end{align}
with pairing
\begin{equation}
    \langle \mathbf{v}^{(1)}, \mathbf{v}^{(2)} \rangle = l^{(1)}_1 n^{(2)}_1 + l^{(2)}_1 n^{(1)}_1 + l^{(1)}_2 n^{(2)}_2 + l^{(2)}_2 n^{(1)}_2 + \sum_{j = 1}^{16} s_j^{(1)}  s_j^{(2)} \, .
\end{equation}
Then, vectors in $\Lambda_N = \Lambda_N^* \cong U \oplus U \oplus \Gamma_{16}$ are characterized by
\begin{align}
    l^{(\ell)}_i, n^{(\ell)}_i \in \bbZ \, , \quad (s_1,...,s_{16})\in \tfrac12 \bbZ \ \ \text{with} \ \ \sum_{j=1}^{16} s_j \in 2\bbZ \, , \ \ s_j - s_k \in \bbZ \ \ \forall j, k \, .
\end{align}

The explicit embedding of the $\fkg$ root lattice $\Lambda_\text{r}^{\fkg}$ into $\Lambda_N$ is given as:
\begingroup\makeatletter\def\f@size{8}\check@mathfonts
\def\maketag@@@#1{\hbox{\m@th\large\normalfont#1}}%
\begin{align}
\left[\begin{array}{cccc|cccccccccccccccc}
1 & 4 & -1&-3 & 0 &-2 &-2 &-1 &-1 &-1 &-1 &-4 & 0 & 0 & 0 & 0 & 0 & 0 & 0 & 0 \\\hline
1 & 2 &-1 &-3 &-1 &-1 &-1 &-1 &-1 &-1 &-1 &-3 & 0 & 0 & 0 & 0 & 0 & 0 & 0 & 0 \\ \hline
0 & 2 & 0 &-2 & 0 &-2 &-1 &-1 &-1 &-1 &-1 &-1 & 0 & 0 & 0 & 0 & 0 & 0 & 0 & 0 \\
0 & 0 & 0 & 0 & 0 & 1 &-1 & 0 & 0 & 0 & 0 & 0 & 0 & 0 & 0 & 0 & 0 & 0 & 0 & 0 \\
0 &-2 & 0 & 1 & 0 & 0 & 1 & 1 & 1 & 1 & 1 & 1 & 0 & 0 & 0 & 0 & 0 & 0 & 0 & 0 \\\hline
0 & 0 & 0 & 0 & 0 & 0 & 0 & 0 & 0 & 1 &-1 & 0 & 0 & 0 & 0 & 0 & 0 & 0 & 0 & 0 \\
0 & 0 & 0 & 0 & 0 & 0 & 0 & 0 & 1 &-1 & 0 & 0 & 0 & 0 & 0 & 0 & 0 & 0 & 0 & 0 \\
0 &-2 & 0 & 1 & 1 & 1 & 1 & 0 & 0 & 1 & 1 & 1 & 0 & 0 & 0 & 0 & 0 & 0 & 0 & 0 \\\hline
1 & 0 & 1 & 0 & 0 & 0 & 0 & 0 & 0 & 0 & 0 & 0 & 0 & 0 & 0 & 0 & 0 & 0 & 0 & 0 \\
0 & 4 &-1 &-3 &-1 &-2 &-2 &-2 &-1 &-1 &-1 &-3 &-1 & 0 & 0 & 0 & 0 & 0 & 0 & 0 \\
0 & 0 & 0 & 0 & 0 & 0 & 0 & 0 & 0 & 0 & 0 & 0 & 1 &-1 & 0 & 0 & 0 & 0 & 0 & 0 \\
0 & 0 & 0 & 0 & 0 & 0 & 0 & 0 & 0 & 0 & 0 & 0 & 0 & 1 &-1 & 0 & 0 & 0 & 0 & 0 \\
0 & 0 & 0 & 0 & 0 & 0 & 0 & 0 & 0 & 0 & 0 & 0 & 0 & 0 & 1 &-1 & 0 & 0 & 0 & 0 \\
0 & 0 & 0 & 0 & 0 & 0 & 0 & 0 & 0 & 0 & 0 & 0 & 0 & 0 & 0 & 1 &-1 & 0 & 0 & 0 \\
0 & 0 & 0 & 0 & 0 & 0 & 0 & 0 & 0 & 0 & 0 & 0 & 0 & 0 & 0 & 0 & 1 &-1 & 0 & 0 \\
0 & 0 & 0 & 0 & 0 & 0 & 0 & 0 & 0 & 0 & 0 & 0 & 0 & 0 & 0 & 0 & 0 & 1 &-1 & 0 \\
0 & 0 & 0 & 0 & 0 & 0 & 0 & 0 & 0 & 0 & 0 & 0 & 0 & 0 & 0 & 0 & 0 & 0 & 1 &-1 \\
0 & 0 & 0 & 0 & 0 & 0 & 0 & 0 & 0 & 0 & 0 & 0 & 0 & 0 & 0 & 0 & 0 & 0 & 1 & 1 
\end{array}\right],
\end{align}\endgroup
whose rows we label by ${\boldsymbol\mu}_1, ..., {\boldsymbol\mu}_{18}$. Here ${\boldsymbol\mu}_1$ and ${\boldsymbol\mu}_2$ are the roots of two $\mathfrak{su}(2)$'s, $({\boldsymbol\mu}_3, {\boldsymbol\mu}_4, {\boldsymbol\mu}_5$ and $({\boldsymbol\mu}_{6}, {\boldsymbol\mu}_7, {\boldsymbol\mu}_{8})$ are the roots of two $\mathfrak{su}(4)$'s. 
$({\boldsymbol\mu}_9, ..., {\boldsymbol\mu}_{18})$ are the roots of $\mathfrak{so}(20)$, with ${\boldsymbol\mu}_{17}, {\boldsymbol\mu}_{18}$ the two branched nodes.
Since these are all ADE-systems, we have ${\boldsymbol\mu}_i = {\boldsymbol\mu}_i^\vee$.

The coweight lattice $\Lambda_\text{cw}^\fkg$ is spanned by the coweights
\begin{equation}
    \overline{\bf w}_i = (C^{-1})_{ij} {\boldsymbol\mu}_j \, , \quad \text{with} \quad C_{ij} = \langle {\boldsymbol\mu}_i , {\boldsymbol\mu}_j \rangle \, .
\end{equation}
Note that $C$ is simply the block-diagonal sum of the Cartan matrices of each simple factor in $\fkg$.
Now we examine the $F$ plane --- the orthogonal subspace to $E := \Lambda_\text{r}^\fkg \otimes \mathbb{R}$, which is two-dimensional in this case. 
Its generators can be chosen to be:
\begin{align}
    \begin{split}
    \boldsymbol\xi_1 &= (-2, 0, 2, 1;0, 0, 0, 0, 0, 0, 0, 2, 0, 0, 0, 0, 0, 0, 0, 0) \, , \\
    \boldsymbol\xi_2 &= (2, 14, -2, -11; -3, -7, -7,  -6, -4, -4, -4, -11, 0, 0, 0, 0, 0, 0, 0, 0) \, ,\\
    \boldsymbol\xi_1^2 &= \boldsymbol\xi_2^2 = -4, \ \ \boldsymbol\xi_1 \cdot \boldsymbol\xi_2 = 0 \, .
    \end{split}
\end{align}

With this basis, a general element $\overline{\bf v}$ of $\Lambda_N^* = \Lambda_N$ can be written as a linear combination of coweights and the $U(1)$ generators:
\begin{equation}
    \overline{\bf v} = (l_1, l_2, n_1, n_1; s_1, \dots, s_{16}) = \sum_{j = 1}^{18} k_j \overline{\bf w}_{j} + m_1 \boldsymbol\xi_1 + m_2 \boldsymbol\xi_2,\ \ \ k_j \in \bbZ \, .
\end{equation}
Modulo the (co-)roots $\Lambda_\text{cr}^\fkg = \Lambda_\text{r}^\fkg$, we find two independent basis vectors of $\Lambda_N \cap E = \Lambda_\text{cc}^G$:
\begin{align}
    \begin{split}
    \hat{\bf c}_1  &= (1, 5, -1, -5; -\tfrac{3}{2}, -\tfrac{5}{2}, -\tfrac{7}{2}, -\tfrac{5}{2}, -\tfrac{3}{2}, -\tfrac{3}{2}, -\tfrac{3}{2}, -\tfrac{9}{2}, \tfrac{1}{2},  \tfrac{1}{2}, \tfrac{1}{2}, \tfrac{1}{2}, \tfrac{1}{2}, \tfrac{1}{2}, \tfrac{1}{2}, -\tfrac{1}{2}) \\
    &= \overline{\bf w}_2 +  \overline{\bf w}_4 + \overline{\bf w}_{17} \, , \\
    \hat{\bf c}_2  &= (1, 1, 0, -1;  \tfrac{1}{2}, -\tfrac{1}{2}, -\tfrac{1}{2}, -\tfrac{1}{2}, \tfrac{1}{2}, -\tfrac{1}{2}, -\tfrac{1}{2}, -\tfrac{3}{2}, -\tfrac{1}{2},  -\tfrac{1}{2}, -\tfrac{1}{2}, -\tfrac{1}{2}, -\tfrac{1}{2}, -\tfrac{1}{2}, -\tfrac{1}{2}, \tfrac{1}{2}) \\
    & = \overline{\bf w}_1 + \overline{\bf w}_7 + \overline{\bf w}_{9} - \overline{\bf w}_{17} \, ,
    \end{split}
\end{align}
each of which defines an order two element, i.e., generates a $\bbZ_2 \subset Z(SU(2)^2 \times SU(4)^2 \times Spin(20)) = \bbZ_2 \times \bbZ_2 \times \bbZ_4 \times \bbZ_4 \times (\bbZ_2 \times \bbZ_2)$, via the the embeddings
\begin{align}
    z(\hat{\bf c}_1) = (0, 1, 2, 0, (1,0)) \, , \quad z(\hat{\bf c}_2) = (1, 0, 0, 2, (0,1)) \, .
\end{align}

The two generators of $\Lambda_N$ that are not within $\Lambda_N \cap E$ are:
\begin{align}
    \begin{split}
    \hat{\bf c}_3 & = (0, 1, 0, -2; 0, -1, -1, -1, 0, -1, -1, -1, 0, 0, 0, 0, 0, 0, 0, 0)= \tfrac{1}{4}\boldsymbol\xi_1 + \overline{\bf w}_2 + \overline{\bf w}_{3}  + \overline{\bf w}_{7} \, , \\
    \hat{\bf c}_4 & = (1, 4, -1,-4; 0, -2, -3, -2, -1, -1, -1, -4, 0, 0, 0, 0, 0, 0, 0, 0)= \tfrac{1}{4}\boldsymbol\xi_2 + \overline{\bf w}_1 + \overline{\bf w}_4 + \overline{\bf w}_{8} \, .
    \end{split}
\end{align}
Their projection under $\pi_E$ onto $\Lambda_\text{cw}^\fkg$ define the following equivalence classes in $Z(\widetilde{G}):$
\begin{equation}
    z(\hat{\bf c}_3) = (0, 1, 1, 2, (0, 0)) \, \quad z(\hat{\bf c}_4) = (1, 0, 2, 1, (0, 0)) \, .
\end{equation}
In summary, we find that the full gauge group is
\begin{equation}
    \frac{[(SU(2)^2 \times SU(4)^2 \times Spin(20))/(\bbZ_2 \times \bbZ_2)] \times U(1)^2}{\bbZ_4 \times \bbZ_4} \, .
\end{equation}

\section{Global Gauge Group of Maximally Enhanced 8d CHL Vacua}
\label{app:big_table}

In this appendix, we present the non-Abelian gauge group $G = \widetilde{G} / {\cal Z}$ of maximally enhanced 8d CHL vacua, i.e., with rank$(G) = 10$.
There are 61 of them \cite{Font:2021uyw,Hamada:2021bbz}, listed in the same order as \cite{Hamada:2021bbz}.
We determined these from their ``parent'' heterotic models, as described in Section \ref{sec:3}.
The global structure of these theories can be obtained with various methods, including that of \cite{Font:2020rsk}.
In practice, we use a generalization of string junctions techniques \cite{Guralnik:2001jh}, which will be elaborated in our upcoming work \cite{junctions_tba}.
There, we will also compute the full global gauge group, including the $U(1)$s.

From the embeddings ${\cal Z} \hookrightarrow Z(\widetilde{G})$, one can explicitly verify that all non-trivial gauge groups are consistent with the vanishing of the mixed 1-form center anomaly \cite{Cvetic:2020kuw}.
We have also checked that the two cases, \#24 and \#52, whose character lattice contains only real representations, satisfy the constraint $\dim(G) + \text{rank}(G) = 0\mod 8$ \cite{Montero:2020icj}:
\begin{equation}
\begin{split}
    \#24:& \quad \text{dim}(Spin(12)) + \text{dim}(Sp (4)) +\text{rank}(Spin(12)) + \text{rank}(Sp (4)) = 112 = 0 \text{ mod } 8 \,, \\
    \#52:& \quad \text{dim}(Spin(16)) + 2 \, \text{dim}(SU (2)) +\text{rank}(Spin(16)) + 2 \, \text{rank}(SU (2)) = 136 = 0 \text{ mod } 8 \,.
\end{split}    
\end{equation}

\renewcommand{\arraystretch}{1.3}
\begin{longtable}{|c|c|c|c|c|}
\caption{All 61 maximally enhanced CHL vacua, together with the simply-connected cover $\widetilde{G} = \prod_i \widetilde{G}_i$ of their non-Abelian gauge group $G = G/{\cal Z}$.
The embedding ${\cal Z} \hookrightarrow Z(\widetilde{G})$ is specified by expressing the generator(s) of ${\cal Z}$ via a tuple $(k_i) \in \prod_i Z(\widetilde{G}_i)$.
If $\widetilde{G}_i = Spin(4n)$, then $k_i = (k_i^{(1)}, k_i^{(2)}) \in Z(Spin(4n)) \cong \bbZ_2 \times \bbZ_2$.
All ADE-factors have Kac-Moody level 2, while the $Sp(n)$ factors have level 1.
Note that, while $Sp(1) \cong SU(2)$ as Lie groups, we will use $Sp(1)$ if the gauge factor is at level 1, and $SU(2)$ if it is at level 2.
\label{tab:big_table}
} \\
\hline
\#  & $\widetilde{G}$    & $\mathcal{Z}$   &  $\mathcal{Z} \hookrightarrow Z(\widetilde{G})$\\ \hline 
1   & $E_8 \times Sp(2)$        &  0   & -  \\  \hline
2   & $E_8 \times Sp(1) \times SU(2)$  &  0   & -   \\  \hline
3   & $E_7 \times Sp(3)$        &  0   & - \\  \hline
4   & $E_7 \times Sp(2) \times SU(2)$  &  $\bbZ_2$  &    $(1, 1, 0)$     \\  \hline
5   & $E_7 \times Sp(1) \times SU(3)$  &  0   & -  \\  \hline
6   & $E_7 \times SU(3) \times SU(2)$  &  $\bbZ_2$  &   $(1, 0, 1)$     \\  \hline
7   & $E_6 \times Sp(4)$        &  0   & -  \\   \hline
8   & $E_6 \times Sp(3) \times SU(2)$  &  0   & -  \\ \hline
9   & $E_6 \times Sp(1) \times SU(4)$  &  0   & -   \\ \hline
10  & $E_6 \times Sp(1) \times SU(3) \times SU(2)$    &  $0$  &  -   \\ \hline
11  & $E_6 \times SU(5)$    &   $0$  &  -  \\ \hline
12  & $Sp(10)$    &   $0$  & -  \\ \hline
13  & $Sp(9) \times SU(2)$    &   $0$  & -  \\ \hline
14  & $Sp(8) \times SU(3)$    &   $\bbZ_2$  &   $(1, 0)$    \\ \hline
15  & $Sp(8) \times SU(2)^2$     &   $\bbZ_2$  &   $(1, 0, 0)$    \\ \hline
16  & $Sp(7) \times SU(3) \times SU(2)$ &  $0$       & -   \\ \hline
17  & $Sp(6) \times SU(5)$    &   $0$  & -  \\ \hline
18  & $Sp(6) \times SU(4) \times SU(2)$    &   $\bbZ_2$  &   $(1, 0, 1)$    \\ \hline
19  & $Sp(6) \times SU(3)^2$     &  $0$      &  -      \\ \hline
20  & $Sp(6) \times SU(3) \times SU(2)^2$  & $\bbZ_2$    &  $(1, 0, 1, 0)$        \\ \hline
21  & $Sp(5) \times Spin(10)$     &  $0$      &  -     \\ \hline
22  & $Sp(5) \times SU(6)$      &  $0$      &  -     \\ \hline
23  & $Sp(5) \times SU(5) \times SU(2)$  &  $0$     &  -     \\ \hline
24  & $Sp(4) \times Spin(12)$     & $\bbZ_2$    &   $(1, (1,1))$        \\ \hline
25  & $Sp(4) \times Spin(10) \times SU(2)$ & $\bbZ_2$    &  $(1, 2, 0)$        \\ \hline
26  & $Sp(4) \times SU(5) \times SU(2)^2$  & $\bbZ_2$    &  $(1, 0, 1, 1)$        \\ \hline
27  & $Sp(4) \times SU(4) \times SU(3) \times SU(2)$   & $\bbZ_2$    &    $(1, 2, 0, 0)$          \\ \hline
28  & $Sp(4) \times SU(3)^2 \times SU(2)^2$   & $\bbZ_2$    &  $(1, 0, 0, 1, 1)$          \\ \hline
29  & $Sp(3) \times SU(7) \times SU(2)$    &  $0$      & -       \\ \hline
30  & $Sp(3) \times SU(6) \times SU(3)$    &  $0$      & -       \\ \hline
31  & $Sp(3) \times SU(5) \times SU(3) \times SU(2)$    &  $0$      & -       \\ \hline
32  & $Sp(3) \times SU(4) \times SU(3)^2$    &  $0$      & -       \\ \hline
33  & $Sp(2) \times Spin(14) \times SU(2)$    &  $\bbZ_2$      &   $(1, 2, 1)$         \\ \hline
34  & $Sp(2) \times Spin(12) \times SU(3)$    &  $\bbZ_2$      &   $(1, (1, 0), 0) $         \\ \hline
35  & $Sp(2) \times Spin(10) \times SU(3) \times SU(2)$    &  $\bbZ_2$      &    $(1, 2, 0, 1)$         \\ \hline
36  & $Sp(2) \times SU(9)$    &  $0$      & -       \\ \hline
37  & $Sp(2) \times SU(7) \times SU(3)$    &  $0$      & -       \\ \hline
38  & $Sp(2) \times SU(6) \times SU(4)$    &  $\bbZ_2$   &   $(1, 3, 0)$         \\ \hline
39  & $Sp(2) \times SU(6) \times SU(2)^3$   &  $\bbZ_2 \times \bbZ_2$ &  \tabincell{l}{$(1, 3, 0, 0, 0)$,\\ $(1, 0, 1, 1, 1)$}          \\ \hline
40  & $Sp(2) \times SU(5)^2$   &  $0$      & -       \\ \hline
41  & $Sp(2) \times SU(5) \times SU(4) \times SU(2)$ & $\bbZ_2$   &    $(1, 0, 2, 1)$          \\ \hline
42  & $Sp(2) \times SU(4)^2 \times SU(2)^2$   &   $\bbZ_2 \times \bbZ_2$ &  \tabincell{l}{$(1, 2, 0, 1, 0)$,\\ $(1, 0, 2, 0, 1)$}        \\ \hline
43  & $Sp(1) \times Spin(18)$    &  $0$      & -       \\ \hline
44  & $Sp(1) \times Spin(10) \times SU(5)$    &  $0$   & -       \\ \hline
45  & $Sp(1) \times SU(10)$       &  $0$      & -       \\ \hline
46  & $Sp(1) \times SU(9) \times SU(2)$       &  $0$      & -       \\ \hline
47  & $Sp(1) \times SU(8) \times SU(2)^2$    & $\bbZ_2$   &  $(0, 4, 1, 1)$          \\ \hline
48  & $Sp(1) \times SU(7) \times SU(3) \times SU(2)$   &  $0$      & -       \\ \hline
49  & $Sp(1) \times SU(6) \times SU(5)$    &  $0$      & -       \\ \hline
50  & $Sp(1) \times SU(6) \times SU(4) \times SU(2)$  &  $\bbZ_2$   &    $(0, 3, 2, 1)$        \\ \hline
51  & $Sp(1) \times SU(5) \times SU(3)^2 \times SU(2)$   &  $0$      & -       \\ \hline
52  & $Spin(16) \times SU(2)^2$   &  $\bbZ_2 \times \bbZ_2$ &   \tabincell{l}{$((1, 1), 1, 1)$,\\ $((0, 1), 0, 0 )$}        \\ \hline
53  & $Spin(12) \times SU(4) \times SU(2)$  &  $\bbZ_2 \times \bbZ_2$ &   \tabincell{l}{$((1, 1), 2, 0)$, \\ $((1, 0), 0, 1)$ }          \\ \hline
54  & $Spin(10)^2$    &     $\bbZ_2$   &  $(2, 2)$          \\ \hline
55  & $SU(10) \times SU(2)$  &     $\bbZ_2$   &  $(5, 1)$         \\ \hline
56  & $SU(8) \times SU(3) \times SU(2)$  &     $\bbZ_2$   &  $(4, 0, 0)$         \\ \hline
57  & $SU(7) \times SU(3)^2$       &  $0$      & -       \\ \hline
58  & $SU(6)^2$  &     $\bbZ_2$   &    $(3, 3)$        \\ \hline
59  & $SU(6) \times SU(5) \times SU(2)$  &     $\bbZ_2$   &   $(3, 0, 1)$          \\ \hline
60  & $SU(6) \times SU(4) \times SU(2)^2$  &  $\bbZ_2 \times \bbZ_2$ &  \tabincell{l}{$(0, 2, 1, 1)$,\\  $(3, 0, 1, 0)$}         \\ \hline
61  & $SU(4)^2 \times SU(3)^2$  &  $\bbZ_2$   &  $(2, 2, 0, 0)$          \\ \hline
\end{longtable}

\end{appendix}

\bibliographystyle{JHEP}
\bibliography{CHL.bib}

\end{document}